\documentclass[aps,prb,twocolumn,unsortedaddress,showpacs]{revtex4}%
\usepackage{graphicx}
\usepackage{epsfig}
\usepackage{amsmath}
\usepackage{amsfonts}
\usepackage{amssymb}

\newcommand{\btau}{\mbox{\boldmath$\tau$}}
\newcommand{\bepsilon}{\mbox{\boldmath$\epsilon$}}
\newcommand{\bnabla}{\mbox{\boldmath$\nabla$}}
\newcommand{\bS}{\mbox{\boldmath$S$}}
\newcommand{\bQ}{\mbox{\boldmath\small$Q$}}
\newcommand{\bq}{\mbox{\boldmath\small$q$}}

\newcommand{\bH}{\mbox{\boldmath$H$}}
\newcommand{\bh}{\mbox{\boldmath$h$}}

\newcommand{\bR}{\mbox{\boldmath\small$R$}}

\newcommand{\ba}{\mbox{\boldmath$a$}}

\newcommand{\br}{\mbox{\boldmath$r$}}
\newcommand{\bu}{\mbox{\boldmath$u$}}
\newcommand{\jlarge}{\mbox{\large$j$}}
\newcommand{\ii}{\mbox{$\rm i$}}

\begin{document}

\title{Heisenberg magnet with modulated exchange.}

\author{I.~A.~Zaliznyak}
\affiliation{Brookhaven National Laboratory, Upton, New York 11973-5000 USA.}

\begin{abstract}
A modification of the ground state of the classical-spin Heisenberg Hamiltonian
in the presence of a weak superstructural distortion of an otherwise Bravais
lattice is examined. It is shown that a slight modulation of the crystal lattice
with wavevector $\bQ_c$ results in a corresponding modulation of the exchange
interaction which, in the leading order, is parametrized by no more than two
constants per bond, and perturbs the spin Hamiltonian by adding the ``Umklapp''
terms $\sim S^\alpha_{\bq} S^\alpha_{\bq \pm \bQ_c}$. As a result, for a general
spin-spiral ground state of the non-perturbed exchange Hamiltonian, an
incommensurate shift of the propagation vector, $\bQ$, and additional new
magnetic Bragg peaks, at $\bQ \pm n\bQ_c$, $n = 1,2,...$, appear, and its energy
is lowered as it adapts to the exchange modulation. Consequently, the lattice
distortion may open a region of stability of the incommensurate spiral phase
which otherwise does not win the competition with the collinear N\'{e}el state.
Such is the case for the frustrated square-lattice antiferromagnet. In addition,
the ``Umklapp'' terms  provide a commensuration mechanism, which may lock the
spin structure to the lattice modulation vector $\bQ_c$, if there is sufficient
easy-axis anisotropy, or a magnetic field in an easy plane.
%and are expected to open gaps in the spin-wave spectrum

\end{abstract}


\maketitle

\section{Introduction}

An interplay between the distortion of a crystal lattice and the magnetic
properties of the material has recently become a subject of renewed interest.
One problem which provides strong motivation for studying the effect of a weak
superstructural modulation on the spin system is that of striped
phases.\cite{TranquadaWakimotoLee,Tranquada1995} These charge-ordered states are
found in lightly doped high-T$_c$ cuprates La$_{2-x}$Sr$_x$CuO$_{4+y}$ (LSCO)
and in related nickelates, and are always associated with a weak superstructural
distortion of the original ``stacked square lattice'' structure of the un-doped
parent material. Incommensurate magnetism in these compounds is usually
interpreted in terms of the segregation of the doped charges into lines which
separate the antiferromagnetic domains (``stripes''), characteristic of the
un-doped material. There is also a modulation of the crystal structure induced
by the charge-stripe segregation, but it is often too small to be observed in
experiment.\cite{TranquadaWakimotoLee} It is clear that the essential effect of
the stripe order on the spin system of cuprates is that of a periodic modulation
of the exchange coupling in the Heisenberg spin Hamiltonian which describes
their magnetic properties.\cite{Coldea2001} However, only the simplest
``average'' consequence of stripe superstructure, in the form of the effective
weakening of the exchange coupling in the direction perpendicular to stripes,
has been considered so far.\cite{CastroNetoHone} A similar problem, of an
interplay between the spin order and a cooperative Jahn-Teller distortion
accompanying the charge order, arises in the context of the charge-ordered
phases in doped manganites.\cite{Orenstein2000}

A number of examples not related to charge ordering but no less interesting,
involve an intriguing interplay between small superlattice modulation and spin
structure in spin-frustrated antiferromagnetic dielectrics. In the
square-lattice antiferromagnet, a distortion could actually be the source of the
frustration. For example, it may generate a well-known generalized Villain
model.\cite{SaslowErwin1992} In the triangular-lattice antiferromagnet (TLA), a
weak distortion may partially release frustration, and result in complicated
spiral phases. Among the simple realizations of the distorted TLA are the
so-called ``row models'' which were extensively studied in the
past.\cite{Zhang,Zhitomirsky1995,Zhitomirsky1996,Trumper1999} However, previous
studies were mainly restricted to a specialized analysis of a few particular
models; no general approach that would allow a unified treatment of the effect
of a small lattice distortion on a spin system has been developed so far.

A number of interesting experimental examples of distorted triangular lattice
antiferromagnets (DTLA), which instigated this study, are found among the
CsNiCl$_3$-type compounds with the general chemical formula ABX$_3$. In the
(anti)ferroelectric phases that are realized in some of these materials at low
temperatures relevant for the magnetic order, the $P6_3/mmc$ hexagonal symmetry
in which they crystallize is lowered, and a fully frustrated triangular lattice
inherent in the original CsNiCl$_3$-type ``stacked triangular lattice'' crystal
structure is slightly distorted. Typically, the high-temperature hexagonal
structure with a Bravais lattice of equivalent magnetic B$^{2+}$ sites is
changed to either a hexagonal $P6_3/cm$ structure with a 3 times larger unit
cell, or to a large-cell orthorhombic structure. These superstructures are
characterized by the appearance of the superlattice Bragg reflections at $\bQ_c
= (\eta,\eta,0)$,\cite{indexing} with $\eta = 1/3$, or $\eta = 1/4,1/8,...$,
respectively. In some cases, as in KNiCl$_3$, both phases are found to coexist
at low temperature. \cite{Petrenko1996} Perhaps, the most intriguing is the case
of RbMnBr$_3$, in which most experiments find the orthorhombic low-$T$
phase,\cite{Zaliznyak2003,Kato,Collins1997} and an incommensurate spiral spin
structure with propagation vector $\bQ = (1/3 + q, 1/3 + q,
1)$,\cite{Zaliznyak2003,Kato,Collins1997,Tikhonov2000} in place of the
commensurate ``triangular'' antiferromagnetic order with $\bQ = (1/3,1/3,1)$,
which is characteristic of the non-distorted hexagonal materials CsMnBr$_3$,
CsNiCl$_3$, \emph{etc}.\cite{Yelon-Cox,Eibshutz,Borovik-Romanov} In a magnetic
field of about 3~T applied in the easy plane the spin structure becomes
commensurate, with $\bQ = (1/8,1/8,1)$. In the similar orthorhombic modification
of KNiCl$_3$, which is a related material but with an easy-axis spin anisotropy,
this latter structure is realized already at $H=0$.

Until now, these experimental findings remained to a large extent unexplained.
One reason for this is that traditionally, the effect of each particular lattice
distortion on the spin Hamiltonian was considered separately, by devising a
specific, generally multi-sublattice spin model (\emph{eg}, the ``row models''),
where the distortion simply defines the particular setup of the near-neighbor
exchange interactions. For long-period structural modulations, this approach
leads to models with a large number of inequivalent spin sites (\emph{eg} up to
8 for $\bQ_c = (1/8,1/8,0)$), resulting in tremendously complicated spin
Hamiltonians, and, therefore, the analysis has never been carried through. The
same problem is outstanding for the stripe phases in LSCO cuprates, where the
most stable superstructure has a pitch of about 1/8 (curiously, it is the same
as that of the antiferroelectric lattice distortion in
RbMnBr$_3$\cite{Zaliznyak2003}). In addition, the modulation has an even longer
period at small doping, and, in general, can also be incommensurate. Here we
devise an alternative approach, which lays grounds for the consistent and
general explanation of spin incommensurability, commensuration transition, and
other phenomena arising from the lattice distortion, that were mentioned above.
We treat the effect of an \emph{arbitrary} but \emph{small} lattice distortion
on the microscopic spin Hamiltonian in the perturbation framework. The analysis
in this paper most directly applies in the case of the dielectrics with
localized spins, although we expect it to hold also for the doped perovskites,
to the extent that the itineracy effects can be neglected.

Consider a system of $N$ equivalent spins on a simple Bravais lattice, coupled
by Heisenberg exchange interactions. The model Hamiltonian, which allows also
for a uniaxial spin anisotropy and a Zeeman energy, is

\begin{equation}
\label{H0}%
{\cal H} = \sum_{i,j} J_{ij}\, \bS_i \cdot \bS_j + D\sum_i \left(S_i^z\right)^2
- \sum_i \bh \cdot \bS_i ,
\end{equation}
where $J_{ij}=J_{ji}$ parameterize the exchange coupling between the spins at
lattice sites $i$ and $j$, $D$ is the anisotropy constant, and $\bh = g\mu_B\bH$
is the magnetic field. Without the anisotropy and magnetic field, the classical
ground state of (\ref{H0}) is a planar spin spiral,
\cite{Yoshimori,Villain,Nagamiya,Lyons-Kaplan,Kaplan} $\bS_j = (S\cos \theta_j,
S\sin\theta_j,0)$, $\theta_j= (\bQ \cdot \br_j)$. The ordering wave vector $\bQ$
corresponds to the minimum of the lattice Fourier transform of the exchange
interaction, $J_{\bq} = \sum_{\br_{ij}} J_{ij} \exp(-\ii \bq \cdot \br_{ij})$,
$\br_{ij} = \br_j - \br_i$ (this includes ferro- and antiferromagnetism,
corresponding to $\bQ = 0$ and, \emph{eg}, $\bQ = (\pi,\pi,\pi)$, respectively).
For non-zero $D$ and $\bh$, the spin structure is, in general, modified by the
appearance of higher-order harmonics in the lattice Fourier transform of the
spin distribution $\bS_{\bq} = \sum_{\br_{j}} \bS_j \exp(-\ii \bq \cdot \br_j)$,
with the wavevectors $\bQ_n = n\bQ$, in addition to
$\bS_{\bQ}$.\cite{CooperElliott,ZaliznyakZhitomirsky,ZhitomirskyZaliznyak} For
small $D$ and $h$ the corresponding terms in the Hamiltonian (\ref{H0}) can be
treated as a perturbation. The perturbation expansion for the correction to the
spiral winding angle has the form  $\delta\theta_j = \sum_n \left( \alpha_n \cos
\left( n{\bQ \cdot \br}_j \right) + \beta_n \sin \left( n{\bQ \cdot \br}_j
\right) \right)$, where the coefficients $\alpha_n$ and $\beta_n$ are of the
order $O( (\frac{|D|}{J})^\frac{n}{2}, (\frac{h}{J})^n
)$.\cite{ZaliznyakZhitomirsky}

In this paper we consider what happens to the classical spiral ground state of
the Hamiltonian (\ref{H0}) if a weak superstructural modulation of the original
Bravais lattice appears in the crystal whose spin system it describes. It is
clear that results obtained for classical spins are subject to corrections from
quantum and thermal fluctuations, and these are often crucial. Nevertheless, if
we find the spiral state that has the lowest energy already on the classical
level, inclusion of the 1/S corrections may still result in this state being the
ground state (clearly, this will always be the case for large enough spin S). In
fact, this may appear to be the case even for small spins, so long as the
spin-wave theory holds. While account for the fluctuations is important, it
falls beyond the scope of this paper and is deferred to further studies.

\section{Superlattice distortion and the spin Hamiltonian} \label{Hamiltonian}

Consider a slight distortion of the crystal structure which is characterized by
the appearance of additional, weak supperlattice Bragg reflections at
wavevectors $\pm \bQ_c$ in the Brillouin zone of the non-distorted Bravais
lattice. Most generally, such a superstructure corresponds to a small harmonic
modulation of the positions, $\br^\mu_{j}$, of the ions and the characteristic
symmetry points of the local electron density distribution (orbitals) in the
lattice,
\begin{equation}
\label{distortion}%
(\br^\mu_j)' = \br^\mu_j + \bepsilon^\mu_j\;,\; \bepsilon^\mu_j =
\bepsilon^\mu_1 \cos(\bQ_c \cdot \br_j^\mu) + \bepsilon^\mu_2 \sin(\bQ_c \cdot
\br_j^\mu).
\end{equation}
Here $j$ numbers the sites of the original Bravais lattice of magnetic
ions,\cite{lattice} and $\mu$ indexes positions of ligands and symmetry points
of the magnetic orbitals within the unit cell. It is shown in Appendix \ref{AA}
that a leading correction to the exchange coupling in this case is most
generally expressed as,
\begin{align}
\label{J_mod}%
J_{ij}' = \tilde{J}_{ij} + \jlarge'_{ij} \cos(\tilde{\bQ_c} \cdot \bR_{ij}) -
\jlarge''_{ij} \sin(\tilde{\bQ_c} \cdot \bR_{ij}),
\end{align}
where $\bR_{ij} = \frac{1}{2} (\bR_i + \bR_j)$ is the middle of the $ij$ bond,
$\jlarge'_{ij} = \jlarge'_{ji},\; \jlarge''_{ij} = \jlarge''_{ji}$ are real and
symmetric, and, most importantly, satisfy all symmetries of the original
lattice which leave $\bQ_c$ and the polarizations
$\bepsilon^\mu_1,\bepsilon^\mu_2$ invariant; the same is true for
$\tilde{J}_{ij}$. In particular, $\tilde{J}_{ij}$, $\jlarge'_{ij}$ and
$\jlarge''_{ij}$ are invariant under the translation group of the non-distorted
lattice. Finally, $\tilde{\bQ_c} = n\bQ_c$, where $n = 1,2$, is the order of the
leading correction to $J_{ij}$ of Eq. (\ref{J_mod}) in terms of the small
parameter $\epsilon \sim \left( \frac{\epsilon^\mu_{1,2} }{r_{ij}} \right) \ll
1$ ($\jlarge'_{ij}, \jlarge''_{ij} \sim \epsilon^n
J_{ij}$).\cite{first_harmonics}

In the first order in $\epsilon$, lattice modulation (\ref{distortion}) results
in the modulation of the exchange coupling $J_{ij}'$ with the same wavevector
$\bQ_c$. It is described by Eq. (\ref{J_mod}), with $\tilde{\bQ_c} = \bQ_c$,
$\tilde{J}_{ij} = J_{ij}$, and $\jlarge'_{ij}, \jlarge''_{ij}$, defined by Eqs.
(\ref{J_mod_12}), (\ref{J_mod_13}) of Appendix \ref{AA}. The structure of the
expressions for $\jlarge'_{ij}, \jlarge''_{ij}$ is quite illuminating. There are
two contributions, one of which is $\sim \sin \left( \frac{\bQ_c \cdot
\br_{ij}}{2} \right)$, and, therefore, antisymmetric with respect to $\bQ_c
\rightarrow -\bQ_c$, and the other depends on the relative alignment of $\bQ_c$
and the distortion polarizations with respect to the bond geometry. Only the
first contribution survives in the simplest case when exchange depends on the
bond length alone, $J_{ij} = J(r_{ij})$; in this case the bonds that are
perpendicular to $\bQ_c$ are not changed by the lattice distortion.

In many important cases the first-order corrections vanish, and the leading
correction in Eq. (\ref{J_mod}) is $\sim \epsilon^2 J_{ij}$, in which case
$\tilde{\bQ_c} = 2\bQ_c$, and $\jlarge'_{ij}, \jlarge''_{ij}$ are given by Eqs.
(\ref{J_mod_7})-(\ref{J_mod_10}). In general, the latter have to be amended in
accordance with (\ref{ligand_modulation}), (\ref{J_mod_11}). Importantly, a
$\bQ_c$-independent contribution which determines the leading correction to the
bond strength $J_{ij}$,
\begin{align}
\label{J_corr}%
\tilde{J}_{ij} = J_{ij} +  \delta J_{ij} \;,\;\; \delta J_{ij} \sim \epsilon^2,
\end{align}
always appears in this order. Because the correction to the bond strength,
$\delta J_{ij}$, arises from replacing $2\cos^2(\bQ_c \cdot \bR_{ij})$,
$2\sin^2(\bQ_c \cdot \bR_{ij})$, with $\cos (2\bQ_c \cdot \bR_{ij}) \pm 1$, it
is directly related with the amplitudes of the exchange modulation. Each
second-order term that contributes to $\jlarge'_{ij}, \jlarge''_{ij}$, also adds
to $\delta J_{ij}$, this is explicit in Eq. (\ref{J_mod_6}). As before, there
are two contributions to $\jlarge'_{ij}, \jlarge''_{ij}$; the one that survives
for $J_{ij} = J(r_{ij})$ does not affect the bonds that are perpendicular to
$\bQ_c$, but is now symmetric with respect to $\bQ_c \rightarrow -\bQ_c$.

Accounting for the corrections to the bond strength, (\ref{J_corr}), is
straightforward and does not require any additional consideration. It simply
amounts to a change in the Fourier-transformed exchange coupling, $J_{\bq}$,
which determines the ground-state energy and the spin-wave spectrum of the
Hamiltonian
(\ref{H0}).\cite{Yoshimori,Villain,Nagamiya,Lyons-Kaplan,Kaplan,CooperElliott,ZaliznyakZhitomirsky}
The significance of this correction, however, is in that a change in $J_{\bq}$
applies directly to the ground state energy, which, therefore, is corrected in
the same order, $O(\epsilon^2)$. As we shall see below, the first-order
contribution to the exchange modulation in Eq. (\ref{J_mod}), $O(\epsilon)$,
corrects the ground state energy of the spin Hamiltonian (\ref{H0}) only in the
second order of perturbation. Therefore, except for special
cases,\cite{small_Q_example} these two contributions have to be treated equally.

To summarize, a superlattice distortion (\ref{distortion}) leads to a harmonic
modulation of the exchange coupling, with either the same wavevector $\bQ_c$, if
it appears as a first-order correction to $J_{ij}$, or with the wavevector
$2\bQ_c$, if it appears in the second order, $\sim \epsilon^2$. There is also a
second-order correction to the bond strength. In the most general case these are
described by Eq. (\ref{J_mod}). In what follows, we study the effect of the
exchange modulation on the ground state of the exchange part of the spin
Hamiltonian (\ref{H0}) (\emph{ie} the case $D=H=0$), which now reads,
\begin{equation}
\label{H_mod}%
{\cal H} = \sum_{i,j} \left( J_{ij} + \jlarge_{ij} {\rm e} ^{\ii  \bQ_c \cdot
\bR_{ij}} + \jlarge^*_{ij} {\rm e} ^{-\ii \bQ_c \cdot \bR_{ij}} \right) \bS_i
\cdot \bS_j \;.
\end{equation}
Here we have introduced a complex $\jlarge_{ij} = \jlarge'_{ij} + \ii
\jlarge''_{ij}$, and omitted the tildes, keeping implicit that $\bQ_c$,
$J_{ij}$, $\jlarge'_{ij}$, $\jlarge''_{ij}$ are all appropriately chosen in
accordance with the situation, as discussed above. While in the absence of the
distortion $J_{ij}$ would satisfy all symmetries of the lattice, exchange
constants in Eq. (\ref{H_mod}) possess only those symmetries of the un-distorted
lattice which preserve $\bQ{_c}$ and the polarizations
$\bepsilon^\mu_1,\bepsilon^\mu_2$ (this includes all translations).

\section{Classical ground state of the modulated-exchange Hamiltonian}

To proceed with finding the ground state of the modulated-exchange Hamiltonian
(\ref{H_mod}), we first rewrite it in terms of the lattice Fourier-transforms,
taking advantage of the lattice translational symmetry,
\begin{align}
\label{H_mod_Fourier}%
\frac{\cal H}{N} &= \sum_{\bq} \{ J_{\bq} \bS_{\bq} \cdot \bS_{-\bq } +
\jlarge_{\bq + \frac{1}{2}\bQ_c} \bS_{\bq} \cdot \bS_{-\bq - \bQ_c} \nonumber \\
 &+ \jlarge^*_{\bq - \frac{1}{2}\bQ_c} \bS_{\bq} \cdot \bS_{-\bq + \bQ_c} \} \;.
\end{align}
Here $\jlarge_{\bq} = \sum_{\br_{ij}} \jlarge_{ij} \exp(-\ii \bq \cdot \br_{ij})
= \jlarge_{-\bq}$, similar with the Fourier-transforms $J_{\bq}$ and $\bS_{\bq}$
introduced above. We note, however, that unlike $J_{\bq}$ which is real,
$\jlarge_{\bq}$ is, in general, complex, so $\jlarge_{\bq}^* \neq
\jlarge_{-\bq}$.

\subsection{General approach}

A general approach to finding the classical ground state for a system of N equal
spins on a simple Bravais lattice that are coupled by isotropic Heisenberg
exchange interaction was developed in Refs.
\onlinecite{Yoshimori,Villain,Nagamiya}, and recently discussed in Ref.
\onlinecite{ZaliznyakZhitomirsky}. We need to solve the mathematical problem of
finding the absolute minimum of a function (\ref{H_mod_Fourier}), which depends
on $N$ classical vector-variables, $\bS_{\bq}$, under $N$ constraints,
\begin{equation}
\label{eq_S}%
\bS_j^2 = S^2 \;, \forall j,
\end{equation}
that are imposed on the length of classical spins. In terms of the
Fourier-components these constraints become,
\begin{equation}
\label{eq_S_Fourier}%
\sum_{\bq'} \bS_{\bq'} \cdot \bS_{\bq - \bq'} = S^2\delta_{\bq,0} \;, \forall
\bq,
\end{equation}
where $\delta_{\bq, \bq'}$ is the 3D Kronekker symbol. Upon introducing $N$
Lagrange multipliers, $\lambda_{\bq}$, a straightforward variation leads to the
following equations for the spin structure that minimizes the Hamiltonian
(\ref{H_mod_Fourier}),
\begin{align}
\label{minimum}%
J_{\bq} \bS_{\bq} + \jlarge_{\bq - \frac{1}{2}\bQ_c} \bS_{\bq
- \bQ_c} + \jlarge^*_{\bq + \frac{1}{2}\bQ_c} \bS_{\bq + \bQ_c} & \nonumber \\
- \sum_{\bq'} \lambda_{\bq'} \bS_{\bq - \bq'} = 0 & ,
\end{align}
$\forall\; \bq$, which have to be solved together with Eq. (\ref{eq_S_Fourier}).
Substituting these into Eq. (\ref{H_mod_Fourier}), we obtain the ground state
(GS) energy per site,
\begin{equation}
\label{E_GS}%
\frac{E_{GS}}{N} = \lambda_0S^2,
\end{equation}
which depends only on $\lambda_{\bq}$ with $\bq = 0$. While this suggests
searching for the solution with $\lambda_{\bq} = \lambda_0 \delta_{\bq,0}$, it
is easily verified by direct substitution that Eq. (\ref{minimum}) does not
allow any non-trivial solutions of this type for $\jlarge_{\bq} \neq 0$. In
fact, using such an ``ansatz'' for $\lambda_{\bq}$ is equivalent to replacing
(\ref{eq_S}) with a single ``weak'' condition, $\sum_j \bS_j^2 = NS^2$. This
``weak'' condition approach, also known as the Luttinger-Tisza method, was
widely employed throughout the early studies of complex spin structures,
\cite{Yoshimori,Villain,Nagamiya,Lyons-Kaplan,Kaplan} as it does lead to the
correct solution in several important particular cases, including the case of a
simple exchange spiral. General reasons for the failure of this approach, which,
in particular, occurs for Hamiltonian (\ref{H0}) with $D \neq 0$, $H \neq 0$,
were discussed in Ref. \onlinecite{ZaliznyakZhitomirsky}. Current situation, in
the form of Eqs. (\ref{H_mod_Fourier}), (\ref{minimum}), presents another
example of such failure.

An alternative approach to finding the ground state and the spin-wave spectrum
of the Hamiltonian (\ref{H0}) for small $D$ and $h$ was used in Refs.
\onlinecite{ZaliznyakZhitomirsky,ZhitomirskyZaliznyak}. It is based on a
perturbative solution for the real-space spin structure that is stationary with
respect to small deviations in the form of a slightly distorted flat spiral. The
correction to the spiral winding angle, $\theta_j$, was obtained in the form of
the expansion, $\delta\theta_j = \sum_n \left( \alpha_n \cos \left( n{\bQ \cdot
\br}_j \right) + \beta_n \sin \left( n{\bQ \cdot \br}_j \right) \right)$, where
the coefficients $\alpha_n$ and $\beta_n$ are of the order
$O((\frac{|D|}{J})^\frac{n}{2}, (\frac{h}{J})^n)$. As a result, additional
harmonics in the Fourier-transform of the spin structure, at wavevectors $\pm
n\bQ$, $n=2,3,...$, and a magnetization, $\bS_0$, appear. In fact, the same
result can be obtained from the conditional minimization of the Hamiltonian
(\ref{H0}) outlined above, if a perturbative solution is searched in the form of
a \emph{harmonic expansion}, $\lambda_{\bq} = \sum_n \lambda_{n}
\delta_{\bq,n\bQ}$, where $\lambda_{n\neq 0} \sim O((\frac{|D|}{J})^\frac{1}{2},
\frac{h}{J}) \cdot \lambda_{|n|-1}$, and $\bS_{\bq} = \sum_n \bS_{n\bQ}
\delta_{\bq,n\bQ}$, $|\bS_{n\bQ}| \sim O(\lambda_{|n|-1})$. It is this approach,
which is both natural, and straightforward to apply to the Hamiltonian
(\ref{H_mod_Fourier}) in order to obtain the spin-wave expansion upon expressing
spins through magnon creation/annihilation operators,\cite{HolsteinPrimakoff}
that we shall employ here.

\subsection{Harmonic expansion for modulated exchange}

Because the modulated-exchange terms in the spin Hamiltonian allow the Umklapp
processes which couple $\bS_{\bq}$ and $\bS_{\bq \pm \bQ_c}$ and, consequently,
couple these Fourier-components in the Eq. (\ref{minimum}), we search for the
solution of the Eqs. (\ref{eq_S_Fourier}), (\ref{minimum}) in the form of the
expansion,
\begin{align}
\label{Sq_expansion}%
\bS_{\bq} = \sum_{n } \left\{ \bS_{\bQ + n\bQ_c} \delta_{\bq, \bQ + n\bQ_c} +
\bS_{\bQ + n\bQ_c}^* \delta_{-\bq, \bQ + n\bQ_c} \right\},
\end{align}
where $S_{\bQ + n\bQ_c} \sim O(\epsilon^n)$. Substituting this into Eq.
(\ref{minimum}) it is easy to see that a non-trivial solution requires that
$\lambda_{\bq} = 0$ unless ${\bq \pm (\bQ + n\bQ_c)} = {\pm (\bQ + n'\bQ_c)}$ is
satisfied for some $n$ and $n'$. Therefore, the general solution for
$\lambda_{\bq}$ can be written as,
\begin{align}
\label{lam_q_expansion}%
\lambda_{\bq} = \sum_{n} \lambda_{n} \delta_{\bq, n\bQ_c},
\end{align}
where $\lambda_{n} \sim O( \epsilon^{|n|} )$. Here and below, if the limits are
not specified explicitly, it is implied that the summation extends over all
integers (this, in particular, implies taking the thermodynamic limit, $N
\rightarrow \infty$).\cite{commensurate_case} With (\ref{Sq_expansion}) and
(\ref{lam_q_expansion}) the energy minimum conditions of Eqs. (\ref{minimum})
become,
\begin{align}
\label{minimum_1}%
& J_{\bQ + n\bQ_c} \bS_{\bQ + n\bQ_c} + \jlarge_{\bQ +(n - \frac{1}{2}) \bQ_c}
\bS_{\bQ + (n-1)\bQ_c} + \nonumber \\
& \jlarge^*_{\bQ +(n + \frac{1}{2}) \bQ_c} \bS_{\bQ + (n+1)\bQ_c} - \sum_{n'}
\lambda_{n'} \bS_{\bQ + (n-n') \bQ_c} = 0,
\end{align}
$\forall\; n$. Subsequently, upon substituting (\ref{Sq_expansion}) into Eq.
(\ref{eq_S_Fourier}), the equal-spin constraint is rewritten as,
\begin{align}
\label{eq_S_1}%
& \sum_{n'} \left( \bS_{\bQ + n'\bQ_c} \cdot \bS_{\bQ + (n-n') \bQ_c} \right) = 0, \\
\label{eq_S_2}%
& 2 \sum_{n'} \left( \bS_{\bQ + (n+n') \bQ_c} \cdot \bS_{\bQ + n'\bQ_c}^*
\right) = S^2 \delta_{n,0},
\end{align}
$\forall\; n$. At this point, (\ref{minimum_1})-(\ref{eq_S_2}) is still a
complicated nonlinear system of equations, and remains such even if we retain
only the terms $\sim O(\epsilon)$ which determine the lowest-order corrections
to the simple exchange spiral.

\subsection{Exchange symmetry}

Further progress into finding the perturbative solution to Eqs.
(\ref{minimum_1})-(\ref{eq_S_2}) which would describe a weakly distorted
exchange spiral, $\{\bS_{\bQ + n\bQ_c}; \lambda_n \}$, is made by employing a
powerful ``exchange symmetry'' argument, which relates back to the Landau theory
of phase transitions. It was developed in Ref. \onlinecite{AndreevMarchenko} as
a basis for the unified Lagrangian description of the long-wavelength,
macroscopic dynamics of the complicated spin systems with complex order
parameters, including spin glasses. Subsequently, this approach was used with
great success, in particular, for calculating the low-energy spin dynamics in a
variety of situations encountered in the non-collinear ground states of
CsNiCl$_3$-type triangular-lattice
antiferromagnets.\cite{Zaliznyak1988,Abarzhi1993,Zhitomirsky1995,Zaliznyak1996}
It is based on a very simple observation, that a macroscopic Lagrangian (or a
Hamiltonian) of a spin system in a state which is described by an order
parameter (at $T \approx 0$), when expressed in terms of the canonical variables
that parameterize the long-wavelength dynamics of this order parameter, has to
satisfy all remaining symmetries of the ground state (order parameter).
Practically, this works as follows. In exchange approximation, possible ground
states are few, easily classified, and Lagrangians are relatively easy to write.
Perturbation account for the anisotropy, magnetic field, \emph{etc}, adds terms
to the Lagrangian (Hamiltonian), which are expansions in powers of the order
parameter, and whose general form is essentially determined by the above
symmetry requirement.

We extend the exchange symmetry argument to the microscopic description of the
present paper by noting that, so long as the solution of Eqs.
(\ref{minimum_1})-(\ref{eq_S_2}) is a weakly distorted simple exchange spiral
$\bS_j = \bS_{\bQ} \exp( {\ii \bQ \cdot \br_j}) + \bS_{\bQ}^* \exp( {- \ii \bQ
\cdot \br_j})$ (\emph{ie} no independent order parameter appears in addition to
$\bS_{\bQ}$), and if the perturbation does not violate the $O(3)$ spin symmetry
of the initial exchange Hamiltonian, all vectors in spin space, $\bS_{\bq}$,
that define the corrections to the initial exchange structure, have to be
proportional to $\bS_{\bQ}$. In other words, the only ``selected'' directions in
spin space which can determine direction of spin vectors in the perturbation
series are those resulting from the spontaneous breaking of the spin symmetry
that already exists in the non-perturbed system, \emph{ie} those defined by
$\bS_{\bQ}$. This holds for the Hamiltonian (\ref{H_mod}), because the
modulated-exchange terms preserve the $O(3)$ spin symmetry. Consequently, we
write,
\begin{align}
\label{S_epsilon}%
\bS_{\bQ + n\bQ_c} = \varepsilon_{n} \bS_{\bQ} \;, \;\; \bS_{-\bQ + n\bQ_c} =
\varepsilon_{-n}^* \bS_{-\bQ},
\end{align}
where $\varepsilon_n \sim O(\epsilon^n)$, and $\varepsilon_0 \equiv 1$. This
simplifies equations (\ref{minimum_1})-(\ref{eq_S_2}) tremendously, as they
shall now involve only scalar variables $\varepsilon_n ,\; \lambda_n$. In
addition, Eq. (\ref{eq_S_1}) is automatically satisfied for $( \bS_{\bQ} )^2 =
0$, which holds in the case of a simple exchange spiral. It requires that spin
vectors of real and imaginary parts of $\bS_{\bQ} = \bS' + \ii \bS''$ are
mutually perpendicular, $(\bS' \cdot \bS'') = 0$, and have equal length. This
length is determined from Eq. (\ref{eq_S_2}),
\begin{equation}
\label{mod_SQ}%
\left| \bS_{\bQ} \right|^2 = \left| \bS' \right|^2 + \left| \bS'' \right|^2 =
\frac{S^2}{2 \sum_{n} \left| \varepsilon_n \right|^2},
\end{equation}
along with the following set of conditions on $\varepsilon_n$,
\begin{align}
\label{eq_S_3}%
\sum_{n'=0}^{n} \left( \varepsilon_{n-n'} \varepsilon_{-n'}^* \right) +
\sum_{n'=1}^{\infty} \left( \varepsilon_{n+n'} \varepsilon_{n'}^* +
\varepsilon_{-n'} \varepsilon_{ -n-n' }^* \right) = 0,
\end{align}
which have to be satisfied $\forall\; n > 0$. The energy minimum conditions of
Eqs. (\ref{minimum}), (\ref{minimum_1}) become, on account of (\ref{S_epsilon}),
\begin{align}
\label{minimum_2}%
& J_{\bQ + n\bQ_c} \varepsilon_{n} - \sum_{n'=0}^n \lambda_{n'}
\varepsilon_{n-n'} \nonumber \\
& + \jlarge_{\bQ +(n - \frac{1}{2}) \bQ_c} \varepsilon_{n-1} +
\jlarge^*_{\bQ + (n + \frac{1}{2}) \bQ_c} \varepsilon_{n+1} \nonumber \\
& - \sum_{n'=1}^\infty \left( \lambda_{n+n'} \varepsilon_{- n'} + \lambda_{n'}^*
\varepsilon_{n+n'} \right) = 0 ,
\end{align}
$\forall\; n \geq 0$. Similar equations for $n < 0$ are solved simultaneously
with  the above, provided that
\begin{equation}
\label{lambda-n}%
\lambda_{-|n|} = \lambda_{|n|}^*,
\end{equation}
in which case they are simply complex conjugates of (\ref{minimum_1}),
(\ref{minimum_2}). We note, that $\bS_{-\bq} = \bS_{\bq}^*$ because $\bS_j$ are
real, so Eqs. (\ref{lambda-n}) just require that Lagrange multipliers used to
account for the conditions (\ref{eq_S}) are also real. On the other hand,
$\varepsilon_{-n} \neq \varepsilon_{n}^*$. The solution $\{ \varepsilon_n;
\lambda_n \}$ of Eqs. (\ref{eq_S_3}), (\ref{minimum_2}) determines the
minimum-energy configuration of the equal-length spins through Eqs.
(\ref{Sq_expansion}), (\ref{S_epsilon}), (\ref{mod_SQ}).

\subsection{Recursion for the perturbation series and the leading-order
solution}

Although they superficially look cumbersome, equations
(\ref{eq_S_3})-(\ref{minimum_2}) are well suited for the perturbation treatment.
Indeed, because
\begin{equation}
\label{orders}%
\lambda_n \bS_{\bQ + n' \bQ_c} \sim O \left( \epsilon^{|n|+|n'|} \right),
\end{equation}
it is easy to see that the first line in Eq. (\ref{minimum_2}) is $\sim
\epsilon^n$, the second line, except for $n=0$, when both terms in it are $\sim
\epsilon^2$, contains a term $\sim \epsilon^n$, and a term $\sim
\epsilon^{n+2}$, and the last line, like the last sum in Eq. (\ref{eq_S_3}),
sums up contributions $\sim \epsilon^{n+2n'}$, $n' = 1,2,...$, \emph{ie} overall
is $O(\epsilon^{n+2})$. Starting with $\varepsilon_0 = 1$, this defines a set of
recursion relations which determine $\{ \varepsilon_n; \lambda_n \}$ for any
given $n$ through their values for $1,2,..., n-1$. In particular, the
leading-order, $\sim O(\epsilon^n)$, condition on $\varepsilon_n$, is,
\begin{align}
\label{epsilon_n}%
\varepsilon_n + \varepsilon_{-n}^* = -\; \sum_{n'=1}^{n-1} \left(
\varepsilon_{n-n'} \varepsilon_{-n'}^* \right) + O(\epsilon^{n+2}) ,
\end{align}
or, explicitly for the first few orders,
\begin{align}
\label{epsilon_1}%
& \varepsilon_1 + \varepsilon_{-1}^* = O(\epsilon^{3}) ,\\
& \varepsilon_2 + \varepsilon_{-2}^* = \varepsilon_1^2 + O(\epsilon^{4}) , \\
& \varepsilon_3 + \varepsilon_{-3}^* = \varepsilon_1 (\varepsilon_2 -
\varepsilon_{-2}^* ) + O(\epsilon^{5}) , \\
& ... \;. \nonumber
\end{align}

The leading-order correction to $\lambda_0$, which determines the ground-state
energy, appears in the second-order perturbation, $\sim \epsilon^2$. In the same
order appears the intensity of the new magnetic Bragg peaks, $\sim |\bS_{\bQ \pm
\bQ_c}|^2$, that are induced by the exchange modulation. Up to this order, we
obtain from Eq. (\ref{minimum_2}),
\begin{align}
\label{2_order}%
& \lambda_0 = J_{\bQ} + \left( \jlarge_{\bQ - \frac{1}{2} \bQ_c} - \lambda_1
\right) \varepsilon_{-1} + \left( \jlarge_{\bQ + \frac{1}{2} \bQ_c}^* -
\lambda_1^* \right) \varepsilon_{1} \nonumber \\
& + O(\epsilon^4) , \\
& \lambda_1 = \jlarge_{\bQ + \frac{1}{2} \bQ_c} + \left( J_{\bQ + \bQ_c} -
\lambda_0 \right) \varepsilon_{1} + O(\epsilon^3) , \\
& \lambda_1^* = \jlarge_{\bQ - \frac{1}{2} \bQ_c}^* + \left( J_{\bQ - \bQ_c} -
\lambda_0 \right) \varepsilon_{-1} + O(\epsilon^3) ,
\end{align}
which, on account of (\ref{epsilon_1}), have the solution,
\begin{equation}
\label{epsilon_solution}%
\varepsilon_1 = \frac{ \jlarge_{\bQ - \frac{1}{2} \bQ_c} - \jlarge_{\bQ +
\frac{1}{2} \bQ_c} }{ J_{\bQ + \bQ_c} + J_{\bQ - \bQ_c} - 2 J_{\bQ } } +
O(\epsilon^3),
\end{equation}
with $\varepsilon_{-1} = - \varepsilon_1^* + O(\epsilon^3)$ (for now, we exclude
from consideration a singular case, $J_{\bQ + \bQ_c} + J_{\bQ - \bQ_c} = 2
J_{\bQ }$). The ground state energy is then determined by
\begin{equation}
\label{lambda_solution}%
\lambda_0 = J_{\bQ} - \frac{ \left| \jlarge_{\bQ - \frac{1}{2} \bQ_c} -
\jlarge_{\bQ + \frac{1}{2} \bQ_c} \right|^2 }{ J_{\bQ + \bQ_c} + J_{\bQ - \bQ_c}
- 2 J_{\bQ } } + O(\epsilon^4) ,
\end{equation}
and, unless $ \jlarge_{\bQ - \frac{1}{2} \bQ_c} = \jlarge_{\bQ + \frac{1}{2}
\bQ_c}$ and the correction vanishes, it is lower than that of the initial,
non-distorted exchange spiral, because $J_{\bQ}$ is a minimum value of
$J_{\bq}$, and, therefore, $J_{\bQ + \bQ_c} + J_{\bQ - \bQ_c} - 2 J_{\bQ } \geq
0$. This is in agreement with a very general argument, that a non-vanishing
second-order perturbation correction always lowers the ground state energy.

\subsection{Some remarks}

It is useful to express the results obtained in the previous section in terms of
the spin-wave spectrum,
\begin{equation}
\label{omega_q}%
\omega_{\bq} = S \sqrt{2(J_{\bq} - J_{\bQ})(J_{\bq + \bQ} + J_{\bq - \bQ} - 2
J_{\bQ})},
\end{equation}
and the $\bq$-dependent transverse (perpendicular to the spin plane) classical
static staggered spin susceptibility,\cite{JensenMackintosh}
\begin{equation}
\label{chi_q}%
\chi_{\perp} (\bq) = \frac{1}{ 2(J_{\bq} - J_{\bQ})},
\end{equation}
of the initial, non-distorted, single-$\bQ$ exchange spiral. The leading new
Fourier-components of the spin density, $\bS_{\bQ \pm \bQ_c}$ of Eqs.
(\ref{S_epsilon}), (\ref{epsilon_solution}), are
\begin{align}
\label{S_Q_solution_1}%
& \bS_{\bQ + \bQ_c} = \left( \frac{ \jlarge_{\bQ - \frac{1}{2} \bQ_c} -
\jlarge_{\bQ + \frac{1}{2} \bQ_c} }{ \chi_\perp ({\bQ_c}) \omega^2_{\bQ_c}
S^{-2} } \right) \bS_{\bQ} + O(\epsilon^3), \\
\label{S_Q_solution_2}%
& \bS_{\bQ - \bQ_c} = \left( \frac{ - \jlarge^*_{\bQ - \frac{1}{2} \bQ_c} +
\jlarge^*_{\bQ + \frac{1}{2} \bQ_c} }{ \chi_\perp ({\bQ_c}) \omega^2_{\bQ_c}
S^{-2} } \right) \bS_{\bQ} + O(\epsilon^3).
\end{align}
It is clear now, that the singular case mentioned in the previous section
corresponds to the exchange modulation with the wavevector at which the
spin-wave energy vanishes. In this case, unless the numerator in Eqs
(\ref{S_Q_solution_1}), (\ref{S_Q_solution_2}) is also zero, the leading
corrections diverge, and the perturbation approach fails. Generally, there are
two soft spots in the spin-wave spectrum of the exchange spiral. They correspond
to the Goldstone modes at $\bq = 0$ and at the magnetic ordering wavevector,
$\bq = \bQ$. For $\bQ_c = \bQ$, the numerator, $( \jlarge_{ \frac{1}{2} \bQ} -
\jlarge_{\frac{3}{2} \bQ} ) $, is, in general, non-zero and the corrections
diverge. Unless $\bq = 0$ is a special (extremum) point of $\jlarge_{\bq}$, the
numerators in Eqs (\ref{S_Q_solution_1}) and (\ref{S_Q_solution_2}) vanish $\sim
Q_c$ in the limit $\bQ_c \rightarrow 0$, while $\omega^2_{\bQ_c}$ in the
denominator is $\lesssim Q_c^2$. Therefore, for a sufficiently long-wavelength
distortion the corrections may become arbitrarily large. It is not at all
unexpected, though, that the perturbation approach fails extrapolation to $\bQ_c
= 0$, where the modulation is absent, and $\bS_{\bQ \pm \bQ_c} \equiv
\bS_{\bQ}$.

Additional soft regions, such as the lines of soft modes, often appear in
frustrated spin systems, due to the accidental cancellation of the interactions.
In such cases spin system is extremely sensitive to structural modulation with
the wavevector that is close to the soft region(s) of the dispersion. The same
is true for distortions that propagate along the direction(s) of weak
interaction (and weak magnon dispersion) in quasi-low-dimensional spin systems.

From Eqs. (\ref{E_GS}) and (\ref{lambda_solution}), the ground state energy is
\begin{align}
\label{E_GS_solution}%
\frac{E_{GS}}{N} = J_{\bQ} S^2 - \frac{ \left| \jlarge_{\bQ - \frac{1}{2} \bQ_c}
- \jlarge_{\bQ + \frac{1}{2} \bQ_c} \right|^2 S^4}{ \chi_\perp ({\bQ_c})
\omega^2_{\bQ_c}} + O(\epsilon^4) .
\end{align}
In the general case, it is lowered in response to the exchange modulation. This
occurs as a result of the appropriate adjustment (bunching) of the initial,
single-$\bQ$ spiral spin structure through the appearance of the additional
Fourier-harmonics, $\bS_{{\bQ} + n \bQ_c }$, $n = \pm 1, \pm 2, ...$. In
addition, the pitch of the primary spiral component, $\bS_{\bQ}$, may also
change, $\bQ \rightarrow \tilde{\bQ}$, because the spiral propagation vector,
$\tilde{\bQ}$, is now defined by the minimum of the corrected energy, Eq.
(\ref{lambda_solution}) or (\ref{E_GS_solution}). In the case where the
modulation of the crystal structure is long-periodic, \emph{ie} for small $Q_c
\ll 1$, and assuming that correction to the propagation vector, $\delta \bQ =
\tilde{\bQ} - \bQ$, is small, $|\delta \bQ| \ll Q$, we can expand
$J_{\tilde{\bQ} \pm \bQ_c}$ and $\jlarge_{\tilde{\bQ} \pm \frac{1}{2} \bQ_c}$ in
Taylor series and obtain,
\begin{align}
\label{small_Qc}%
& \lambda_0 = J_{\bQ} + \frac{1}{2} \left( \delta \bQ \cdot \bnabla_{} \right)^2
J_{\bQ} \nonumber \\
& - \frac{3 \left| \left(\bQ_c \cdot \bnabla_{} \right) \jlarge_{\bQ}
\right|^2}{ \left( \delta \bQ \cdot \bnabla_{} \right) \left[ \left(\bQ_c \cdot
\bnabla_{} \right)^2 J_{\bQ} \right] + 3 \left(\bQ_c \cdot \bnabla_{} \right)^2
J_{\bQ} } + O(\epsilon^3) ,
\end{align}
where $\bnabla = \frac{\partial}{\partial \bq}$. Note, that unlike Eqs
(\ref{S_Q_solution_1}), (\ref{S_Q_solution_2}) this, generally, does not diverge
for $\bQ_c \rightarrow 0$. If, in the correction term, we cancel $Q_c^2$ and
then expand the denominator keeping only the leading term in $\delta Q$, Eq.
\ref{small_Qc} takes the form $\lambda_0 = J_{\bQ} + c_0 + c_1 (\delta Q) + c_2
(\delta Q)^2$, with $c_1 = O(\epsilon^2)$ and $c_2 = O(\epsilon)$. Clearly,
minimum of this expression occurs, in the general case, for non-zero $\delta Q =
O(\epsilon)$. However, if the linear term $ \sim \delta Q$ in the denominator of
Eq. \ref{small_Qc} vanishes, then $\delta \bQ = 0$. Also, in many important
cases such as, for example, the nearest-neighbor non-frustrated antiferromagnet,
$\bQ$ is a special symmetry point of $ \jlarge_{\bQ}$ ($ \jlarge_{\bQ} \sim
\epsilon J_{\bQ}$), and the correction term identically vanishes by symmetry.
Therefore, simple exchange structures, such as antiferromagnets, are usually
insensitive to small long-periodic modulations of the exchange coupling in the
spin Hamiltonian (\ref{H_mod}).

\section{Some examples}

Now we shall apply the formalism developed above to several representative one-
and two-dimensional (1D and 2D) systems. The fact that the ordered mean-field
(MF) ground state, such as analyzed in this paper, in low dimension is unstable
against the fluctuations, \cite{Landau1937} will not be a concern here. First, a
single-$\bQ_c$ structural distortion considered in this paper is homogeneous
within the crystallographic planes that are perpendicular to $\bQ_c$. In many
cases these planes contain one or two unit cell directions, so the modulation
only exists along the remaining direction(s), and the distortion is explicitly
2D, or 1D, respectively (\emph{cf}, Fig. 1). The bonds that are not changed by
the distortion cancel out in the resulting expressions for the corrections to
the spin structure and to the ground state energy, Eqs.
(\ref{epsilon_solution}), (\ref{lambda_solution}). Therefore, even though the
Hamiltonian (\ref{H0}) may be on a 3D lattice, the distortion corrections will
be the same as for the lower-dimensional system. Also, in many quasi-low-D
materials the essential physics is 1D or 2D, while the MF order is stabilized by
weak interaction in the remaining direction(s). Such is the case of high-$T_c$
cuprates which are made of two-dimensional square-lattice layers. In the
presence of a charge order the layers are modulated, as illustrated in Fig. 1
(b). In fact, the MF analysis has certain value even for purely low-D systems,
because it highlights possible phases and instabilities, and guides the behavior
of the critical points, at least asymptotic ($S \rightarrow \infty$).

%***************************Figure 1*****************************************
\begin{figure}[b]
\vspace{-0.2in}
\begin{center}
\label{examples_of_modulated_J}%
\includegraphics[height=6.4 in]%width=3.2in, ]
{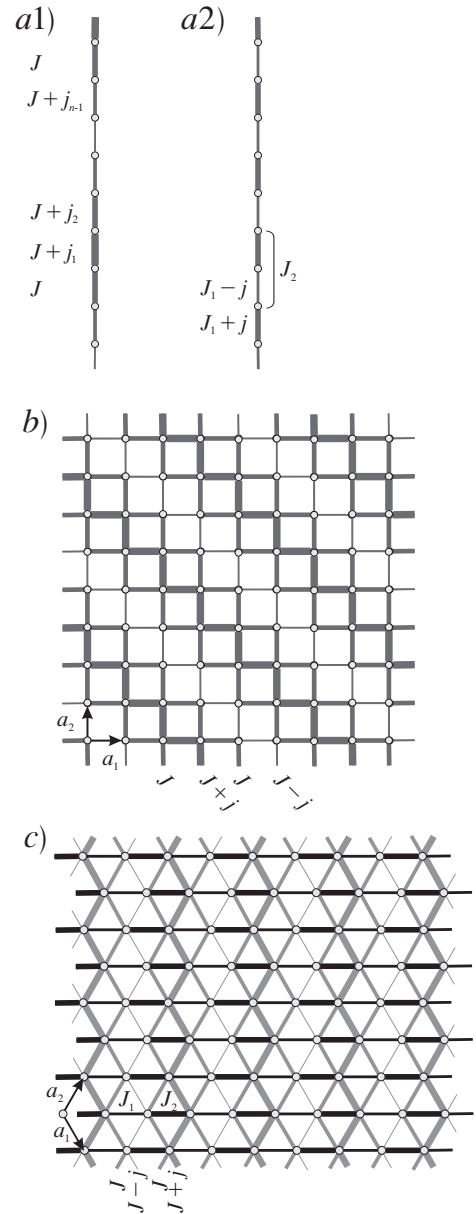}%
\caption{Examples of the modulated exchange patterns discussed in the text.
Different bond thicknesses illustrate different coupling strengths. (a1)
n-merized 1D chain with modulated nearest-neighbor coupling, $Q_c =
\frac{2\pi}{n}$; (a2) dimerized (n=2) antiferromagnetic chain with frustrating
next-nearest neighbor coupling, $4 J_2 \geq J_1$. (b) "Stripes" on a square
lattice with diagonal modulation, $Q_c = (\frac{2\pi}{n}, \frac{2\pi}{n})$;
$n=4$ case is shown. (c) Generalized "staggered row model" on a triangular
lattice, obtained with $Q_c = (\frac{2\pi}{n}, \frac{2\pi}{n})$. Shown is the
case of $n=4$; $n$ equals 8 in the practical case of RbMnBr$_3$. }
\end{center}
\vspace{-.2in}
\end{figure}
%***************************Figure 1*****************************************

\subsection{n-merized 1D antiferromagnetic chain}

As a simple example, consider the antiferromagnetic spin chain with $n$-periodic
nearest-neighbor exchange coupling (ferromagnetic case, $J <0$, is trivial,
because small bond modulation has no effect on the ground state). It corresponds
to a modulation with $Q_c = \frac{2\pi}{n}$, Fig. 1(a1). In this case, $J_{\bq}
= 2J \cos{q}, \; \jlarge_{\bq} = 2\jlarge \cos{q}$, the modified GS energy is
determined by
\begin{align}
\label{lambda0_a1}%
\lambda_0 = 2J \cos{Q} \left( 1 + \left| \jlarge/J \right|^2 \tan^2 Q \right),
\end{align}
and does not depend on $n$. Clearly, there are new local extrema at $\cos^2{Q}
\approx \left|\jlarge/J \right|^2$, corresponding to a spiral with pitch of
$\approx \pi/2$, which appear on account of the exchange modulation. However,
the new minimum is very shallow, $\lambda_0 \approx -4 |\jlarge|$, much higher
in energy than the global minimum, $\lambda_0 = -2 J$, which remains at $Q=\pi$.

A more interesting 1D example is illustrated in Fig. 1(a2), where in addition to
the nearest-neighbor coupling, $J_1$, there is a frustrating
next-nearest-neighbor exchange, $J_2$. In this case $J_{\bq} = 2J_1 \cos{q} +
2J_2 \cos{2q}$ and, for $4 J_2 \geq J_1$, frustration (for classical spins)
results in a spiral MF ground state, with a pitch defined from the condition
$\cos{Q} = -\frac{J_1}{4J_2}$. For the sake of simplicity we consider only
modulation of the nearest-neighbor coupling, \emph{ie} $\jlarge_2 = 0$. In this
case, as before, $\jlarge_{\bq} = 2\jlarge_1 \cos{q}$, and,
\begin{align}
\label{lambda0_a2}%
\lambda_0 = J_{Q} + \frac{2 \left| \jlarge_1 \right|^2 \sin^2 Q}{J_1 \cos{Q} + 4
J_2 \cos^2{\frac{\pi}{n}} \cos{2Q}}.
\end{align}
The minimum of this expression is achieved for $\tilde{Q} = Q + \delta Q$ which
differs from $Q = \cos^{-1} \left( -\frac{J_1}{4J_2} \right)$. For small $\delta
Q$ the leading correction is,
\begin{align}
\label{delta_a2}%
\delta Q = \left| \frac{\jlarge_1}{2J_2} \right|^2 \frac{\cos^2{Q} -
\cos{\frac{2\pi}{n}}}{\tan{Q} (1 - \cos{\frac{2\pi}{n}} \cos{2Q})^2},
\end{align}
and diverges for $Q \rightarrow \pi$, \emph{ie} for frustrated 1D
antiferromagnet with $J_2 = 0.25J_1$. Perhaps, this interesting finding is an
indication of an instability towards a spontaneous $n$-merization and a spin gap
formation in the vicinity of this point. In fact, for S=1/2 quantum spin chain
the whole region $J_2 > J_{2c} \approx 0.25 J_1$ is believed to belong to a
spin-gap phase,\cite{Allen1995,Haldane1982,White1996} including the special
point, $J_2 = 0.5J_1$, where the spontaneously dimerized ground state is known
exactly.\cite{MajumdarGhosh} While the most recent numerical estimate for S=1/2
chain is $J_{2c} \approx 0.24 J_1$,\cite{White1996} a $J_{2c} = 0.25 J_1$ for $S
\rightarrow \infty$ coincides with the earlier result obtained from the
semi-classical mapping on the nonlinear sigma-model.\cite{Allen1995} Clearly,
studying the susceptibility of spin system towards bond modulation is a proper
way to investigate its instability towards $n$-merization (plaquette formation)
and to characterize the corresponding phase diagram. To this end it seems
possible to develop a perturbation approach similar to the one presented here
and starting with the Hamiltonian (\ref{H_mod}) also for quantum spins, but this
goes far beyond the scope of this paper.

\subsection{Square lattice antiferromagnet with diagonal modulation}

A square-lattice, nearest-neighbor antiferromagnet with diagonal modulation
corresponds to $\bQ_c = (\frac{2\pi}{n}, \frac{2\pi}{n})$, Fig. 1(b), and may be
of direct relevance for the charge-ordered stripe phases in doped LSCO cuprates
and related perovskites. In the isotropic case the bond strengths and the
modulation amplitudes are equal in two directions, $J_{\bq} = 2J \left(\cos
\left( \bq \cdot \ba_1 \right) + \cos \left( \bq \cdot \ba_2\right) \right)$,
$\jlarge_{\bq} = \epsilon J_{\bq}$, and, upon switching to $Q = \bq \cdot
\frac{(\ba_1 + \ba_2)}{2}$ and $Q' = \bq \cdot \frac{(\ba_1 - \ba_2)}{2}$, the
problem is factorized and reduced to the one-dimensional one considered above.
The distortion-corrected spin GS energy is determined by $\lambda_0$ of Eq.
(\ref{lambda0_a1}), which is simply multiplied by $\cos Q'$. As before, we find
that nearest-neighbor antiferromagnetism is stable with respect to the bond
modulation. In fact, the same conclusion is reached even if the amplitude of the
bond modulation is different in two directions, so that $\jlarge_{\bq} =
2\jlarge_1 \cos \left( \bq\cdot \ba_1 \right) + 2\jlarge_2 \cos\left( \bq \cdot
\ba_2 \right)$.

Again, an interesting situation occurs if there is frustration. In the case of
the square-lattice antiferromagnet it is introduced by the diagonal,
next-nearest-neighbor coupling, $J' > 0$. This model has been extensively
studied in recent years,
\cite{ChandraDoucout,Chakravarty1989,Lieb1999,Singh1999,Sushkov,Tchernyshyov2003}
since it was predicted that in a region of parameters in the vicinity of
$J/(2J') = 1$ it has a disordered, spin-liquid ground
state.\cite{ChandraDoucout,Chakravarty1989} For S=1/2 quantum spins this was
proposed as a possible candidate for the resonating-valence-bond
state.\cite{Anderson1987,Kivelson1987} In this case, $J_{\bq} = 4J \cos{Q}
\cos{Q'} + 2J'( \cos{2Q} + \cos{2Q'} ) $, and, if both side and diagonal bonds
are modulated, $\jlarge_{\bq} = 4\jlarge \cos{Q} \cos{Q'} + 2\jlarge'( \cos{2Q}
+ \cos{2Q'} ) $. Consequently, we obtain from Eq. (\ref{lambda_solution}),
\begin{align}
\label{lambda0_b}%
\lambda_0 = J_{\bQ} + \frac{4 \sin^2 Q \left| \jlarge \cos{Q'} + 2\jlarge' \cos
{\frac{\pi}{n}} \cos{Q} \right|^2}{J \cos{Q} \cos{Q'} + 2 J'
\cos^2{\frac{\pi}{n}} \cos{2Q}}.
\end{align}

In the absence of bond modulation the ground state is determined by the
hierarchy of the local minima of $J_{\bQ}$. It depends on the relative strength
of the nearest-neighbor coupling $J$, and the next-nearest-neighbor, diagonal
coupling $J'$, which is parameterized by $\alpha = \frac{J}{2J'}$. For weak
frustration, $\alpha > 1$, the global minimum is that with $\sin{Q} = \sin{Q'} =
0$. It corresponds to the conventional, collinear N\'{e}el antiferromagnetic
order with a single propagation vector $\bQ = (\pi,\pi)$, and the ground state
energy is $\frac{1}{N}E_{(\pi,\pi)} = -4J S^2 (1 - \frac{1}{2\alpha})$. Although
there are four equivalent $\bQ$-points in the BZ, $(\pm \pi, \pm \pi)$, $(\mp
\pi, \pm \pi)$, which restore the lattice $C_4$ rotational symmetry, they are
related through addition of the appropriate reciprocal lattice vectors $\btau$,
so there is no true GS degeneracy in the $\bQ$-space. The only GS degeneracy is
the rotational symmetry in spin space which corresponds to the $O(3)$ symmetry
of the Heisenberg spin Hamiltonian.

For strong frustration, $\alpha \leq 1$, additional ground state degeneracy
occurs on the MF level. For $\alpha < 1$ there are two non-equivalent
lowest-energy minima of $J_{\bQ}$, satisfying $\cos{Q} = \cos{Q'} = 0$. They
correspond to two pairs of equivalent $\bQ$-points in the BZ, $(\pm \pi, 0)$ and
$(0, \pm \pi)$, which represent the antiferromagnetic order propagating along
$x$ and $y$ axis, respectively. The GS energy is $\frac{1}{N}E_{(\pi,0)} = -4J'
S^2 = -4J S^2 \frac{1}{2\alpha}$. This double degeneracy in $\bQ$-space can be
used to construct a continuum of states which are the linear combinations of the
above two. This continuous GS degeneracy is usually described in terms of two
decoupled $\sqrt{2} \times \sqrt{2}$ sublattices based on the diagonals of the
original square lattice, which is transparent in the limit $J' \gg J$. Each
sublattice has an antiferromagnetic order, but there may be an arbitrary angle
between the two, because the mean field from one sublattice cancels on the sites
of the other. This continuous degeneracy is lifted by zero-point or thermal
fluctuations which prefer collinear arrangements of the two sublattices in the
GS. This is a famous example of the order by disorder phenomenon in a frustrated
magnet. \cite{Shender1982,Henley1989}

The most interesting situation occurs for $\alpha = 1$, when, on the MF level,
there is a continuous GS degeneracy even in the $\bQ$-space. The minimum
condition for $J_{\bQ}$ becomes $\cos{Q} = \cos{Q'}$, and is satisfied for any
spiral with the propagation vector $\bQ$ that belongs to the square with the
vertices at $(\pm \pi, \pm \pi)$, $(\mp \pi, \pm \pi)$. All these states belong
to the global minimum, and have the same energy, $\frac{1}{N}E_{\alpha=1} = - 2J
S^2 = -4J' S^2$. It is this continuous $\bQ$-space degeneracy which is at the
origin of the spin-liquid phase conjectured for the values of $\alpha$ in a
finite region around the special point $\alpha = 1$.
\cite{ChandraDoucout,Chakravarty1989,Lieb1999,Singh1999,Sushkov}

Importantly, for $\alpha \leq 1$ the spiral states with $\bQ \approx (\pi, \pi)$
are in close competition with the collinear states. In particular, consider an
extremum of $J_{\bQ}$ which is a local minimum along the diagonal direction,
parallel to the lattice modulation wavevector $\bQ_c$, whose energy in the
absence of the modulation is $\frac{1}{N}E_{Q} = - 2\alpha J S^2$. It
corresponds to a spiral with the propagation vector defined by $ Q' = 0$, $
\cos{Q} = - \frac{J}{2J'}$, \emph{ie} $\bQ = \left( \cos^{-1}( -\frac{J}{2J'}),
\cos^{-1}( -\frac{J}{2J'}) \right)$ (there is also a degenerate state with $\bQ$
at $90^\circ$, respecting the $Q \leftrightarrow Q'$ symmetry of the square
lattice). Except for $\alpha = 1$ the energy of this extremum is higher than
that for the decoupled antiferromagnetic sublattices, $E_{(\pi,0)}$. It is
clear, however, from Eq. (\ref{lambda0_b}) that, while the energy of the
collinear antiferromagnetic states is insensitive to bond modulation, the energy
of the spiral GS can be lowered as it adapts to distortion! Therefore, at least
on the MF level, spiral may become the lowest energy state (\emph{ie} the ground
state) for some range of the parameter $\alpha$ in the vicinity of 1 (whose
width is $\sim O( \epsilon^2)$). For the long-periodic modulations, $Q_c \ll 1$,
and for $\jlarge' = 0$, it is easy to find that spiral phase is stable for $1 -
| \jlarge /J |^2 \leq \alpha < 1$. The ``nominal'' spiral propagation vector
$\bQ$ is obtained by minimizing Eq. (\ref{lambda0_b}), similarly with the case
of the frustrated 1D chain. Again, our finding clearly indicates the instability
of the frustrated square-lattice antiferromagnet with $J/(2J')$ close to 1 with
respect to the bond-modulated states. By selecting spiral spin GS the ``order by
distortion'' mechanism proposed here competes with the ``order by disorder''
phenomenon, which prefers collinear states. At least for large enough S spiral
always wins in some vicinity of $\alpha = 1$.

\subsection{Generalized row models on triangular lattice}

Because of inherent frustration, the triangular lattice antiferromagnet, Fig. 1
(c), is a very interesting case to consider. Without modulation, a minimum of
$J_{\bq} = 2J \left( \cos \left( \bq \cdot \ba_1 \right) + \cos \left( \bq \cdot
\ba_2 \right) + \cos \left( \bq\cdot( \ba_1 + \ba_2)\right)\right)$ is achieved
for a commensurate spiral GS with propagation vector $\bQ = (\frac{2\pi}{3},
\frac{2\pi}{3})$. It corresponds to a non-collinear GS structure where spins are
aligned along one of the 3 directions, at 120$^\circ$ degrees with each other.
Structural distortion may result in a variety of coupling patterns where
equivalent bonds are related by translations perpendicular to $\bQ_c$. These can
be classified as ``generalized row models'', the simplest of which is the
original row model of Ref. \onlinecite{Zhang}. It can arise, for example, as a
second-order effect from the distortion with $\bQ_c = (\pm \pi, \pi)$. Because
$2\bQ_c = \btau$ ($\btau$ is a reciprocal lattice vector), it is equivalent to a
homogeneous uniaxial compression, \emph{ie} it simply changes $J_{\bq}$ and does
not result in a bond modulation. Nevertheless, the ground state becomes
incommensurate, with pitch of the spiral determined by the coupling anisotropy.
\cite{Zhang,Zhitomirsky1995,Zhitomirsky1996}

A general modulation of the triangular lattice with $\bQ_c = (\frac{2\pi}{n}, -
\frac{2\pi}{n})$ results in a ``staggered row model'', where the horizontal rows
have equal coupling. On the other hand, modulation with $\bQ_c =
(\frac{2\pi}{n}, \frac{2\pi}{n})$ leads to a ``zigzag row model'', an example of
which with $n=4$ is shown in Fig. 1(c), and the one with $n=8$ is relevant for
the phases realized in RbMnBr$_3$ and
KNiCl$_3$.\cite{Petrenko1996,Kato,Zaliznyak2003} In this case $\jlarge_{\bq} =
4\jlarge \cos{Q} \cos{Q'} + 2\jlarge \cos{2Q} $ and it is easy to see that the
energy of the modulated state is given by the same expression as for the
frustrated square lattice, Eq. (\ref{lambda0_b}), but with $\jlarge' = \jlarge$
and $J' = J$ (as in the previous example, we use $Q = \bq \cdot \frac{(\ba_1 +
\ba_2)}{2}$ and $Q' = \bq \cdot \frac{(\ba_1 - \ba_2)}{2}$). Exchange modulation
leads to a deviation from the commensurate 120$^\circ$ triangular spin
structure. The leading correction to the propagation vector is determined from
$\cos \tilde{Q} = - \frac{1}{2} + \delta$, where
\begin{align}
\label{delta_c}%
\delta  = \left| \frac{\jlarge}{J} \right|^2 \frac{ \left( \cos{\frac{\pi}{n}} -
\cos{\frac{2\pi}{n}} \right) \left(\frac{5}{4} - \frac{9}{4} \cos{\frac{\pi}{n}}
+ \frac{5}{2} \cos^2{\frac{\pi}{n}} \right) }{2 \left( \frac{1}{2} + \cos^2
{\frac{\pi}{n}} \right)^2}.
\end{align}
In the two limiting cases, $n=2$ and $n \gg 1$, we find $\delta = \frac{5}{2}
\left| \frac{\jlarge}{J} \right|^2 $ and $\delta \approx \left( \frac{\pi}{2n}
\right)^2 \left| \frac{\jlarge}{J} \right|^2$, respectively. It is clear that
$\delta$ is small for all $n$, so there is no evidence for an instability
towards $n$-merization in the case of ideal triangular lattice. Perhaps, such
evidence can be found in the anisotropic, quasi-1D case, where exchange in
zigzag rows is much smaller than that in straight rows, or vice versa.

\section{Summary and conclusions}

In a great variety of important practical cases the complex crystal structure
which is at the origin of the intricate magnetic behavior in magnetic material
results from a small superstructural distortion of a much simpler structure, in
which the magnetic ions form a primitive Bravais lattice. Reduction of the
crystal symmetry related to the appearance of even a single, commensurate with
the original lattice, superstructural Bragg reflection at a wavevector $\bQ_c$,
formally requires folding the original Brillouin zone to a much smaller one,
many times reducing its volume. Consequently, the magnetic system is usually
described in terms of multiple spin sites, and  multiple sheets of spin
excitations. Not only does this greatly complicate understanding and predicting
magnetic properties, such violent modification of the $\bq$-space clearly seems
an unsatisfactory way to account for a small distortion of the crystal
structure. Moreover, BZ folding is not an option for the incommensurate
structural modulations, such as arise in various charge-density-wave ordered
states.

In this paper the effect of a small lattice modulation with single propagation
vector $\bQ_c$ on the system of localized spins, coupled by Heisenberg exchange
interaction (\ref{H0}), was considered. It was found that lattice distortion
results in a modulation of the exchange coupling which, to the leading order, is
parameterized by no more than two constants per bond, Eq. (\ref{H_mod}) (this is
valid irrespective of whether the spins are quantum or classical). There are
also corrections of the order $O(\epsilon^2)$ ($\epsilon$ is a small parameter
that parameterizes the lattice distortion) to the exchange constants $J_{ij}$ in
the covariant part of the Hamiltonian. It should be noted here that, although
the distortion considered is small, the resulting corrections to the coupling
constants need not be small compared to the initial values of the couplings in
the spin Hamiltonian, which may be small themselves. Therefore, the result
expressed by Eq. (\ref{H_mod}) is quite general and does not automatically imply
the condition $ |\jlarge_{ij}| \ll |J_{ij}|$. In other words, the ``Umklapp''
terms $\sim S^\alpha_{\bq} S^\alpha_{\bq \pm \bQ_c}$ added to the spin
Hamiltonian by small lattice distortion may be relatively large, even larger
than the original exchange interactions.

While it would be interesting to study the modulated-exchange Hamiltonian for
quantum spins and for the arbitrary values of $ |\jlarge_{ij} / J_{ij}|$, it is
a formidable task which is beyond the scope of this paper. Here we developed a
perturbative scheme for finding the mean field ground state of the Hamiltonian
(\ref{H_mod}) which is valid for classical spins, $S \gg 1$, and in the case of
small exchange modulation, $ |\jlarge_{ij} / J_{ij}| \sim \epsilon \ll 1$. One
of the initial motivations here was to develop a basis for the spin-wave theory
in weakly distorted crystal structures in terms of the modification of the
magnon spectrum in the original, large Brillouin zone of the non-distorted
Bravais lattice. The other, no less important motivation, was to see whether it
would be possible to understand, already on the mean field level, the
incommensurate phases observed in the distorted triangular-lattice
antiferromagnets RbMnBr$_3$ and KNiCl$_3$, and in the doped, distorted
square-lattice antiferromagnets, such as LSCO or related Ni, Mn and Co
materials, such as La$_{1.5}$Sr$_{0.5}$CoO$_4$. \cite{Zaliznyak2000}

The essential results of this paper are expressed by Eqs.
(\ref{epsilon_solution}), (\ref{lambda_solution}), or, equivalently, by Eqs.
(\ref{S_Q_solution_1}) - (\ref{E_GS_solution}). They show, that a transverse,
equal-spin spiral structure, which is the ground state of the initial Heisenberg
Hamiltonian, adapts to the exchange modulation through appearance of the
additional Fourier-harmonics, $\bS_{{\bQ} + n \bQ_c }$, $n = \pm 1, \pm 2, ...$
(bunching). As a result, in a general case the GS energy is lowered by exchange
modulation. In addition, the pitch of the primary spiral component, $\bS_{\bQ}$,
may also change, $\bQ \rightarrow \tilde{\bQ}$, because it is now defined by the
minimum of modulation-corrected energy, Eq. (\ref{E_GS_solution}).

Applying these results to several particular examples of the topical frustrated
spin systems appears quite revealing. We find that in the case of the frustrated
square-lattice antiferromagnet with diagonal coupling $J'$, such that $\alpha =
J/(2J')$ is close to 1, lattice modulation opens a region of stability of the
incommensurate spiral phase. This ``order by distortion'' phenomenon
\cite{Tchernyshyov2002} competes with ``order by disorder'', which prefers
collinear arrangements of two antiferromagnetic sublattices. The incommensurate
spiral phase with the propagation vector $\tilde{\bQ} = (\pi \pm \delta, \pi \pm
\delta)$ close to $(\pi, \pi)$ wins for the range $O(\epsilon^2)$ of the
parameter $\alpha$ in the vicinity of $\alpha = 1$. This provides a plausible
explanation for the incommensurate spin order observed in
La$_{1.5}$Sr$_{0.5}$CoO$_4$ \cite{Zaliznyak2000} and in a number of other doped
perovskites, and may also be of a direct relevance for the doped LSCO materials.
It is important to mention here that incommensurate spin-ordered phases are
among the most interesting and puzzling features of doped layered perovskites.
In the absence of distortion one needs at least a third-neighbor coupling in
order to stabilize spiral MF ground state for the Heisenberg spin Hamiltonian on
square lattice.

Exchange modulation in Heisenberg antiferromagnet on a distorted triangular
lattice leads to an incommensurate shift of the spiral propagation vector, in
qualitative agreement with what is observed in RbMnBr$_3$. However, Eq.
(\ref{delta_c}) implies that $\tilde{\bQ}$ is decreased compared to $\bQ =
(\frac{2\pi}{3}, \frac{2\pi}{3})$ for the ideal triangular lattice, while
$\tilde{\bQ} \approx 2\pi \cdot(0.357,0.357)$ is observed in experiments.
\cite{Zaliznyak2003,Kato} Therefore, it is likely that the shift in RbMnBr$_3$
is mainly due to the anisotropic corrections to the nearest-neighbor coupling,
which are captured already in the simplest row model.
\cite{Zhitomirsky1995,Zhitomirsky1996} Nevertheless, correction of Eq.
(\ref{delta_c}) is not unimportant. In the case of $n=4$, which may be relevant
for RbMnBr$_3$, it gives the same magnitude shift of the ordering wavevector,
$\delta Q$, as measured in experiment, for $\left| \jlarge/J \right|^2 \approx
0.2$ (for $n=8$, $\left|\jlarge/J \right|^2 \approx 1$ is needed).

More importantly, bunching of the spin spiral as a result of the lattice
distortion provides, already on the mean field level, a possible explanation for
the commensuration transition in RbMnBr$_3$ and for the long-periodic
lattice-commensurate structure in the related phase of KNiCl$_3$. Indeed, an
easy-axis anisotropy, such as in KNiCl$_3$, or a magnetic field applied within
the easy plane, as in RbMnBr$_3$, also lead to bunching of the exchange spiral,
generating additional Fourier-harmonics, $\bS_{\bq}$, at $\bq = n \tilde{\bQ}$,
$n = \pm 2, \pm 3, ...$. \cite{ZaliznyakZhitomirsky} Appearance of these
Fourier-components in the spin distribution lowers the spin-anisotropy and the
Zeeman energy, but competes with the modulated exchange, which requires
additional Fourier-components at $\bq = \tilde{\bQ} + n \bQ_c$. Therefore, for
some finite value of the easy-axis anisotropy, or the in-plane magnetic field, a
commensuration transition may be expected, where $\bQ$ becomes equal to $m
\bQ_c$ with some integer $m$. In the lattice-commensurate phase both sets of
additional harmonics coincide, and both the modulated exchange energy, and the
spin-anisotropy and Zeeman energy, can be lowered simultaneously (this, of
course, should off-set the increase in the un-modulated exchange energy caused
by the shift in $\bQ$). Extending the results of this paper and those of Ref.
\onlinecite{ZaliznyakZhitomirsky} to the Hamiltonian (\ref{H0}) with modulated
exchange of the Eq. (\ref{H_mod}) and $D,H \neq 0$ in order to map out such
phase diagram is one of the most obvious directions for further studies.

\begin{acknowledgments}
It is a pleasure to thank F. Essler, S. Maslov and M. Zhitomirsky for
encouraging discussions, and to acknowledge the financial support by the DOE
under Contract \#DE-AC02-98CH10886.
%and by the Theory Institute at Brookhaven National Laboratory.
\end{acknowledgments}

\appendix

\section{Effect of lattice modulation on exchange coupling}
\label{AA}%

The simplest assumption, and the one which is most often employed in literature,
\cite{Pytte} is that the exchange coupling $J_{ij}$ between the magnetic ions
(spins) at positions $\br_i$ and $\br_j$ only depends on the distance between
the sites, $J_{ij} = J(|\br_{ij}|)$. However, in magnetic dielectrics this
coupling most often results from the superexchange and, therefore, generally
also depends on the position(s) of ligand ions which bridge the superexchange
path,
\begin{equation}
\label{J_dependence}%
J_{ij} = J(|\br_{ij}|, \left\{ \br^{\mu_{ij}} \right\}) = J(r_{ij}, \left\{
\bu^{\mu_{ij}} \right\}).
\end{equation}
Here $\mu_{ij}$ numbers the ligands that participate in the $ij$ bond and
$\br^{\mu_{ij}}$ are their positions which are most naturally parameterized in
terms of the offsets, $\bu^{\mu_{ij}} = \br^{\mu_{ij}} - \bR_{ij}$, from the
bond center, $\bR_{ij} = \frac{1}{2} (\bR_i + \bR_j)$.

In addition, the superexchange coupling may also depend on the angles of orbital
overlaps. This dependence can be parameterized in terms of the positions of some
particular symmetry points in the local electron density distribution, and
accounted for in (\ref{J_dependence}) by including these points among $\left\{
\br^{\mu_{ij}} \right\}$. While these additional degrees of freedom do lift some
non-essential symmetries which are present in the particular case when the
exchange coupling only depends on the bond length, $J_{ij} = J(|\br_{ij}|)$,
they do not change the general structure of the corrections to the exchange
coupling resulting from the lattice modulation which are summarized by Eq.
(\ref{J_mod}) in the main text. In what follows, we first discuss the particular
case of $J_{ij} = J(|\br_{ij}|)$, and then the general case of Eq.
(\ref{J_dependence}).

\subsection{Modulation of the bond length only}

First, consider the effect of displacement of the magnetic ions alone. In
presence of the superstructural modulation (\ref{distortion}), the bond lengths
become $r_{ij}' = |\br_{ij} + \bepsilon_{ij}|$, with $\bepsilon_{ij} =
\bepsilon_j-\bepsilon_i$ given by
\begin{align}
\label{bond_modulation}%
\bepsilon_{ij} = 2\sin \left( \frac{\bQ_c\cdot \br_{ij}}{2} \right) \left(
-\bepsilon_1 \sin(\bQ_c \cdot \bR_{ij}) + \bepsilon_2 \cos(\bQ_c \cdot
\bR_{ij})\right),
\end{align}
where, as usual, $\br_{ij} = \br_j - \br_i$ and $\bR_{ij} = \frac{1}{2} (\bR_i +
\bR_j)$.

Expanding the exchange coupling, $J_{ij}' = J(r_{ij}')$, modified by the
distortion (\ref{bond_modulation}) in a Taylor series in small displacement,
$\epsilon_{ij} \ll r_{ij}$,
\begin{equation}
\label{J_mod_1}%
J_{ij}' = J_{ij} + \sum_n \frac{1}{n!} \left(\bepsilon_{ij} \cdot
\frac{\partial}{\partial \br_{ij}}\right)^n J(r_{ij}),
\end{equation}
we find, up to a second order in $\left( \frac{\epsilon_{ij}}{r_{ij}} \right)
\sim \epsilon \ll 1$,
\begin{align}
\label{J_mod_2}%
J_{ij}' = & J_{ij} + \frac{(\bepsilon_{ij}\cdot \br_{ij})}{r_{ij}}
\frac{\partial J(r_{ij})}{\partial r_{ij}} + \frac{1}{2} \frac{(\bepsilon_{ij}
\cdot \br_{ij})^2}{r_{ij}^2}\frac{\partial^2 J(r_{ij})}{\partial
r_{ij}^2} \nonumber \\
+ & \frac{1}{2}\left( \frac{\bepsilon_{ij}^2}{r_{ij}} -
\frac{(\bepsilon_{ij}\cdot \br_{ij})^2}{r_{ij}^3} \right) \frac{\partial
J(r_{ij})}{\partial r_{ij}} + O(\epsilon^3).
\end{align}
For  $(\bepsilon_{ij} \cdot \br_{ij}) \neq 0$ and $\frac{\partial
J(r_{ij})}{\partial r_{ij}} \neq 0$ the leading contribution is given by the
first-order term, $\sim \epsilon$, and we obtain,
\begin{align}
\label{J_mod_3}%
J_{ij}' = J_{ij} + \jlarge'_{ij} \cos(\bQ_c \cdot \bR_{ij}) - \jlarge''_{ij}
\sin (\bQ_c\cdot \bR_{ij}) + O(\epsilon^2),
\end{align}
where
\begin{align}
\label{J_mod_4}%
\jlarge'_{ij} = & 2\sin \left( \frac{\bQ_c\cdot \br_{ij}}{2} \right)
\frac{\partial
J(r_{ij})}{r_{ij} \partial r_{ij}} (\bepsilon_2\cdot \br_{ij}), \\
\label{J_mod_5}%
\jlarge''_{ij} = & 2\sin \left( \frac{\bQ_c\cdot \br_{ij}}{2} \right)
\frac{\partial J(r_{ij})}{r_{ij} \partial r_{ij}} (\bepsilon_1\cdot \br_{ij}).
\end{align}
Clearly, to the first order, there is no $\bQ_c$-independent correction which
would change the bond strength $J_{ij}$. Such a correction does appear in the
second order in $\epsilon$.

If $(\bepsilon_{ij} \cdot \br_{ij}) = 0$, \emph{ie} the displacements are
perpendicular to the bonds, the leading correction is given by the second-order
term $\sim \bepsilon_{ij}^2$. In fact, this important situation is often
encountered in practice, in particular, it is the basis for the so-called ``row
models'' for triangular lattice antiferromagnets. These are typically thought to
be realized through a modulation of the hexagonal lattice where the
displacements of the magnetic ions are parallel to the $C_6$ symmetry axis, and
are perpendicular to the bonds in the hexagonal plane in which the modulation
propagates. RbMnBr$_3$ is believed to present an example of such situation. For
such case of a transverse structural modulation we obtain,
\begin{align}
\label{J_mod_6}%
J_{ij}' = & J_{ij} + \tilde{\jlarge'_{ij}} \cos(2 \bQ_c \cdot \bR_{ij}) -
\tilde{\jlarge''_{ij}} \sin(2 \bQ_c \cdot \bR_{ij}) \nonumber \\
+ & s_{ij} \left( (\tilde{\jlarge'_{ij}})^2 + (\tilde{\jlarge''_{ij}})^2
\right)^{\frac{1}{2}} + O((\bepsilon_{ij} \cdot \br_{ij})^2, \epsilon^3),
\end{align}
where $s_{ij} = {\rm sign} \left( \frac{\partial J(r_{ij})}{\partial r_{ij}}
\right)$, and,
\begin{align}
\label{J_mod_7}%
\tilde{\jlarge'_{ij}} = & \sin^2 \left( \frac{\bQ_c\cdot \br_{ij}}{2} \right)
\frac{\partial J(r_{ij})}{r_{ij} \partial r_{ij}}
\left(\bepsilon_{2}^2 - \bepsilon_{1}^2 \right), \\
\label{J_mod_8}%
\tilde{\jlarge''_{ij}} = & \sin^2 \left( \frac{\bQ_c\cdot \br_{ij}}{2} \right)
\frac{\partial J(r_{ij})}{r_{ij} \partial r_{ij}} 2 \left(\bepsilon_{1}\cdot
\bepsilon_{2} \right).
\end{align}
The first-order correction to $J'_{ij}$ also vanish if $\frac{\partial
J(r_{ij})}{\partial r_{ij}} = 0$. In this case the leading contribution comes
from the last second-order term in Eq. (\ref{J_mod_2}), and is also expressed by
Eq. (\ref{J_mod_6}), but with  $s_{ij} = {\rm sign} \left( \frac{\partial^2
J(r_{ij})}{\partial r_{ij}^2} \right)$, and

\begin{align}
\label{J_mod_9}%
\tilde{\jlarge'_{ij}} = & \sin^2 \left( \frac{\bQ_c\cdot \br_{ij}}{2} \right)
\frac{\partial^2 J(r_{ij})}{r_{ij}^2 \partial r_{ij}^2}
\left((\bepsilon_{2}\cdot \br_{ij})^2 - (\bepsilon_{1}\cdot \br_{ij})^2 \right), \\
\label{J_mod_10}%
\tilde{\jlarge''_{ij}} = & \sin^2 \left( \frac{\bQ_c\cdot \br_{ij}}{2} \right)
\frac{\partial^2 J(r_{ij})}{r_{ij}^2 \partial r_{ij}^2} 2 (\bepsilon_{1}\cdot
\br_{ij}) (\bepsilon_{2}\cdot \br_{ij}),
\end{align}
in place of (\ref{J_mod_7}),(\ref{J_mod_8}).

A $\bQ_c$-independent correction which changes the bond strength $J_{ij}$ first
appears in the second order, Eq. (\ref{J_mod_6}). In general, it is obtained by
summing up all $\bQ_c$-independent contributions from all second-order terms in
(\ref{J_mod_2}). Their common multiplier, $\sin^2 \left( \frac{\bQ_c\cdot
\br_{ij}}{2} \right)$, makes the structure of this correction rather simple. It
does not affect the bonds that are perpendicular to the direction of propagation
of the lattice distortion, while those bonds that are symmetric with respect to
this direction are modified equally. On the other hand, $\jlarge'_{ji}$ and
$\jlarge''_{ij}$ of Eqs. (\ref{J_mod_4}), (\ref{J_mod_5}) which describe the
first-order correction to the exchange coupling resulting from the lattice
modulation on are $\sim \sin \left( \frac{\bQ_c\cdot \br_{ij}}{2} \right)$, and,
therefore, are antisymmetric with respect to $\bQ \rightarrow -\bQ$.

\subsection{Modulation of the ligand positions, etc}

It is also straightforward to account for the dependence of the superexchange
coupling, Eq. (\ref{J_dependence}), on the positions, $\bu^{\mu_{ij}}$, of the
ligand ions and the symmetry points of the local electron density distribution
which define the orbital overlaps. In the presence of the superstructural
distortion (\ref{distortion}),
\begin{align}
\label{general_modulation}%
(\br^{\mu_{ij}})' = \br^{\mu_{ij}} + \bepsilon^{\mu}_1 \cos(\bQ_c \cdot
\br^{\mu_{ij}}) + \bepsilon^{\mu}_2 \sin(\bQ_c \cdot \br^{\mu_{ij}}),
\end{align}
where $\mu$ indexes different types of $\br^{\mu_{ij}}$ positions within the
unit cell, and the polarization vectors $\bepsilon^{\mu}_{1,2}$ which
parameterize the displacement for point of type $\mu$ are determined by the
superlattice Bragg intensities that appear with distortion. This can be
rewritten as,
\begin{align}
\label{ligand_modulation}%
(\bu^{\mu_{ij}})' = \bu^{\mu_{ij}} + \bepsilon^{\mu_{ij}}_1 \cos(\bQ_c \cdot
\bR_{ij}) + \bepsilon^{\mu_{ij}}_2 \sin(\bQ_c \cdot \bR_{ij}),
\end{align}
where the new polarization vectors, $\bepsilon^{\mu_{ij}}_{1,2}$ now depend on
$\bQ_c$. They are obtained by rotating $\bepsilon^{\mu}_{1,2}$ through an angle
$\phi_{\mu_{ij}} = (\bQ_c \cdot \bu^{\mu_{ij}})$, and subtracting
$\bepsilon_{1,2} \cos (\frac{1}{2} \bQ_c \cdot \br_{ij})$ (this accounts for
change in the bond center position $\bR_{ij}$), correspondingly.

Consequently, in the general case of Eq. (\ref{J_dependence}), the Taylor series
(\ref{J_mod_1}) for $J_{ij}'$ has to be amended, by adding,
\begin{equation}
\label{J_mod_11}%
J_{ij}' \rightarrow J_{ij}' + \sum_n \frac{1}{n!} \sum_{\mu_{ij}}
\left(\bepsilon^{\mu_{ij}} \cdot \frac{\partial}{\partial
\bu^{\mu_{ij}}}\right)^n J \left(r_{ij}, \left\{ \bu^{\mu_{ij}} \right\}
\right),
\end{equation}
whose  first and second-order terms are easily rewritten in the form of Eq.
(\ref{J_mod_3}) and Eq. (\ref{J_mod_6}), respectively. Therefore, the account
for the modulation of the positions $\bu^{\mu_{ij}}$ in the general expression
for the superexchange, (\ref{J_dependence}), simply amounts to amending the
coefficients $\jlarge'_{ij}$ and $\jlarge''_{ij}$ in equations (\ref{J_mod_3})
and (\ref{J_mod_6}), in accordance with Eqs. (\ref{ligand_modulation}),
(\ref{J_mod_11}). For example, additional first-order terms which appear in
(\ref{J_mod_11}) change the expressions of Eqs. (\ref{J_mod_4}) and
(\ref{J_mod_5}) as follows,
\begin{align}
\label{J_mod_12}%
\jlarge'_{ij} \rightarrow \jlarge'_{ij}
%= \;& 2\sin \left( \frac{\bQ_c\br_{ij}}{2} \right) \frac{\partial
%J(r_{ij})}{r_{ij} \partial r_{ij}} (\bepsilon_2\br_{ij}) \nonumber \\
&+ \sum_{\mu_{ij}} \left(\bepsilon^{\mu_{ij}}_1 \cdot \frac{\partial}{\partial
\bu^{\mu_{ij}}}\right) J \left(r_{ij}, \left\{ \bu^{\mu_{ij}} \right\} \right), \\
\label{J_mod_13}%
\jlarge''_{ij} \rightarrow \jlarge'_{ij}
%= \;& 2\sin \left( \frac{\bQ_c\br_{ij}}{2} \right) \frac{\partial
%J(r_{ij})}{r_{ij} \partial r_{ij}} (\bepsilon_1\br_{ij}) \nonumber \\
& -\sum_{\mu_{ij}} \left(\bepsilon^{\mu_{ij}}_2 \cdot \frac{\partial}{\partial
\bu^{\mu_{ij}}}\right) J \left(r_{ij}, \left\{ \bu^{\mu_{ij}} \right\} \right).
\end{align}
Using Eqs. (\ref{ligand_modulation}), (\ref{J_mod_11}) it is easy to write out
similar expressions for the coefficients of the second-order contribution of Eq.
(\ref{J_mod_6}).

Clearly, a number of symmetry properties of the coefficients $\jlarge'_{ij}$ and
$\jlarge''_{ij}$ given by Eqs. (\ref{J_mod_4})-(\ref{J_mod_10}) which are
present for $J_{ij} = J(r_{ij})$, disappear upon account for the modulation of
the positions $\bu^{\mu_{ij}}$. In particular, for the first-order corrections
to vanish, not only should the displacements of the lattice sites (magnetic
ions) be perpendicular to the bonds, but all of the displacements
$\bepsilon^{\mu_{ij}}_{1,2}$ should be perpendicular to the corresponding
gradients of $J_{ij} = J(r_{ij}, \left\{ \bu^{\mu_{ij}} \right\})$ with respect
to $\bu^{\mu_{ij}}$. However, it is clear that, because $\jlarge'_{ij}$,
$\jlarge''_{ij}$ and, in general, $\tilde{J_{ij}}$, are functions on the
non-distorted lattice which also depend on the modulation wavevector $\bQ_c$ and
the polarizations $\bepsilon^{\mu_{ij}}_{1,2}$, they are invariant with respect
to all symmetry operations of that initial lattice which do not change $\bQ_c$
and $\bepsilon^{\mu_{ij}}_{1,2}$. Importantly, this includes the translation
group of the non-distorted lattice, which means that the new couplings and the
exchange modulation amplitudes which are related by that lattice translations
are equal.


\begin{thebibliography}{99}                                                                                                %

\bibitem{TranquadaWakimotoLee}
J.~M.~Tranquada, % {\it et~al.}
J.~D.~Axe, N.~Ichikawa, A.R.~Moodenbaugh, Y.~Nakamura and S.~Uchida
 Phys.~Rev.Lett. {\bf 78}, 338(1997);
 Phys.~Rev.~B {\bf 59}, 14712 (1999).
% S.~Wakimoto {\it et~al.},
 %R.~J.~Birgeneau, M.~A.~Kastner, Y.~S.~Lee, R.~Erwin, P.~M.~Gehring,
 %S.~H.~Lee, M.~Fujta, K.~Yamada, Y.~Endoh, K.~Hirota, G.~Shirane
% Phys.~Rev.~B {\bf 61}, 3699 (2000);
% Y.~S.~Lee {\it et~al.},
 %R.~J.~Birgeneau, M.~A.~Kastner, Y.~Endoh, S.~Wakimoto, K.~Yamada,
 %R.~Erwin, S.~H.~Lee, G.~Shirane
% Phys.~Rev.~B {\bf 60}, 3643 (1999).

\bibitem{Tranquada1995}
J.~M.~Tranquada, % \emph{et al}
B.~J.~Sternlieb, J.~D.~Axe, N. Nakamura, S.~Uchida, Nature {\bf 375}, 561,
(1995).

\bibitem{Coldea2001}
R. Coldea, %\emph{et al}
S. M. Hayden, G. Aeppli, T. G. Perring, C. D. Frost, T. E. Mason, S.-W. Cheong,
Z. Fisk Phys. Rev. Lett. {\bf 86}, 5377 (2001).

\bibitem{CastroNetoHone}
A.~H.~Castro-Neto and D.~Hone, Phys. Rev. Lett. {\bf 76}, 2165 (1996).

\bibitem{Orenstein2000}
J.~Orenstein and A.~J.~Millis, Science {\bf 288}, 468 (2000).

\bibitem{SaslowErwin1992}
W.~M.~Saslow, R.~Erwin, Phys. Rev. B {\bf 45}, 4759 (1992).

\bibitem{Zhang}
W.-M.~Zhang, W.~M.~Saslow, M.~Gabay, M.~Benakli, Phys. Rev. B {\bf 48}, 10204
(1993); W.-M.~Zhang, W.~M.~Saslow, M.~Gabay, Phys. Rev. B {\bf 44}, 5129 (1991).

\bibitem{Zhitomirsky1995}
M.~E.~Zhitomirsky, O.~A.~Petrenko, and L.~A.~Prozorova, Phys. Rev. B {\bf 52},
3511 (1995).

\bibitem{Zhitomirsky1996}
M.~E.~Zhitomirsky, Phys. Rev. B {\bf 54}, 353 (1996).

\bibitem{Trumper1999}
A.~E.~Trumper, Phys. Rev. B {\bf 60}, 2987 (1999).

\bibitem{indexing}
Throughout this paper, we index the wavevectors in the Brillouin zone of the
original, non-distorted structure, where magnetic ions occupy the sites of a
Bravais lattice.

\bibitem{Petrenko1996}
O.~A.~Petrenko, M.~A.~Lumsden, M.~D.~Lumsden, M.~F.~Collins, J.~Phys.: Condens.
Matter {\bf 8}, 10899 (1996); O.~A.~Petrenko, M.~F.~Collins, C.~V.~Stager,
B.~F.~Collier, Z.~F.~Tun, J.~Appl. Phys. {\bf 79}, 6614 (1996).

\bibitem{Zaliznyak2003}
I.~A.~Zaliznyak, \emph{et al}, unpublished (2003).

\bibitem{Kato}
T.~Kato, J.~Phys.~Soc.~Jpn. {\bf 71}, 300 (2001); T.~Kato, T.~Asano, Y.~Ajiro,
S.~Kawano, T.~Ishii, K.~Iio, Physica B {\bf 213}\&{\bf 214}, 182 (1995);
T.~Kato, K.~Machida, T.~Ishii, K.~Iio, T.~Mitsui, Phys. Rev. B {\bf 50}, 13039
(1994) ; T.~Kato, T.~Ishii, Y.~Ajiro, T.~Asano, S.~Kawano J.~Phys.~Soc.~Jpn.
{\bf 62}, 3384 (1993).

\bibitem{Collins1997}
M.~F.~Collins, and O.~A.~Petrenko, Can. J.~Phys. {\bf 75}, 605 (1997).

\bibitem{Tikhonov2000}
A.~M.~Tikhonov, S.~V.~Petrov, Phys. Rev. B {\bf 61}, 9629 (2000).

\bibitem{Yelon-Cox}
W.~B.~Yelon and D.~E.~Cox, Phys.~Rev.~B {\bf 6}, 204 (1972); Phys.~Rev.~B {\bf
7}, 2024 (1973).

\bibitem{Eibshutz}
M.~Eibshutz, R.~C.~Sherwood, F.~S.~L.~Hsu, and D.~E.~Cox, AIP Conf. Proc. 17,
864 (1972).

\bibitem{Borovik-Romanov}
A.~S.~Borovik-Romanov, S.~V.~Petrov, A.~M.~Tihkonov, B.~S.~Dumesh, JETP Letters
{\bf 64}, 225 (1996); JETP {\bf 86}, 197 (1998).

\bibitem{Yoshimori}
A.~Yoshimori, J.~Phys.~Soc.~Jpn. {\bf 14}, 807 (1959).

\bibitem{Villain}
J.~Villain, J.~Phys.~Chem.~Solids {\bf 11}, 303 (1959).

\bibitem{Nagamiya}
T.~Nagamiya, in {\it Solid State Physics\/}, edited by F.~Seitz, D.~Turnbull and
H.~Ehrenreich, Vol.~20, 305 (Academic Press, New York, 1967); T.~Nagamiya,
T.~Nagata, and Y.~Kitano, Progr. Teor. Phys. {\bf 27}, 1253 (1962).

\bibitem{Lyons-Kaplan}
D.~H.~Lyons and T.~A.~Kaplan, Phys. Rev. {\bf 120}, 1580 (1960).

\bibitem{Kaplan}
T.~A.~Kaplan, Phys. Rev. {\bf 124}, 329 (1961).

\bibitem{CooperElliott}
B.~R.~Cooper, R.~J.~Elliott, S.~J.~Nettel, and H.~Suhl, Phys. Rev. {\bf 127}, 57
(1962); B.~R.~Cooper and R.~J.~Elliott, Phys. Rev. {\bf 131}, 1043 (1963).

\bibitem{ZaliznyakZhitomirsky}
I. A. Zaliznyak and M. E. Zhitomirsky, cond-mat/0306370; JETP {\bf 81(3)}, 579
(Zh. Eksp. Teor. Fiz. {\bf 108}, 1052) (1995).

\bibitem{ZhitomirskyZaliznyak}
M. E. Zhitomirsky and I. A. Zaliznyak, Phys. Rev. B {\bf 53}, 3428 (1996).

\bibitem{lattice}
We consider materials where in the absence of a distortion magnetic ions occupy
the sites of a simple Bravais lattice. In addition, a number of the non-magnetic
ligand ions, inherent in the chemical formula, may also be present in the unit
cell basis of the crystal structure.

\bibitem{first_harmonics}
Such simple expression with a single-$\bQ$ modulation is only true for the first
two orders in $\epsilon$. In general, the $n$-th order correction is $\sim
\left( \bepsilon_1 \cos(\bQ_c \cdot\br_{ij}) + \bepsilon_2 \sin(\bQ_c
\cdot\br_{ij}) \right)^n$. While it invariably includes modulation with the
wavevector $n\bQ_c$, it may also include lower harmonics. For example, both
modulations with $3\bQ_c$ and $\bQ_c$ appear in the third order.

\bibitem{small_Q_example}
For example, if either of the contributions vanishes by lattice and/or
distortion symmetry.

\bibitem{HolsteinPrimakoff}
T. Holstein, H. Primakoff, Phys. Rev. {\bf 58}, 1098 (1940).

\bibitem{commensurate_case}
In the case when the modulation is commensurate, \emph{ie} there is a number $m$
such that $m \bQ_c = \btau$ ($\btau$ is a reciprocal lattice vector), the sums
are limited to $n \leq m$, and the minimum conditions are finite systems of $m$
equations. However, as far as the leading corrections are concerned, for large
$m>2$ this is no different from the general incommensurate case.

\bibitem{AndreevMarchenko}
A.~F.~Andreev and V.~I.~Marchenko, Usp. Fiz. Nauk {\bf 130}, 39 (1980)
[Sov.~Phys. Uspekhi {\bf 23}, 21 (1980)].

\bibitem{Zaliznyak1988}
I.~A.~Zaliznyak, V.~I.~Marchenko, S.~V.~Petrov, L.~A.~Prozorova., and
A.~V.~Chu\-bukov, JETP Lett. {\bf 47}, 211 (1988).

\bibitem{Abarzhi1993}
S.~I.~Abarzhi, M.~E.~Zhitomirsky, O.~A.~Petrenko, S.~V.~Petrov, and
L.~A.~Prozorova, Zh. Eksp. Teor. Fiz. {\bf 104}, 3232 (1993) [JETP {\bf 77}, 521
(1993)].

\bibitem{Zaliznyak1996}
I.~A.~Zaliznyak, N.~N.~Zorin, S.~V.~Petrov, JETP Lett. {\bf 64}, 473 (1996).

\bibitem{JensenMackintosh}
J.~Jensen and A.~R.~Mackintosh, "Rare Earth Magnetism", Clarendon Press, Oxford
(1991).

\bibitem{Landau1937}
L.~D.~Landau, Zh. Exp. and Teor. Fiz. {\bf 7}, 627 (1937).

\bibitem{Allen1995}
D.~Allen and D.~S\'{e}n\'{e}chal, Phys.~Rev.~B {\bf 51}, 6394 (1995);
\emph{ibid}  {\bf 55}, 299 (1997).

\bibitem{Haldane1982}
F.~D.~M.~Haldane, Phys.~Rev.~B {\bf 25}, 4925 (1982).

\bibitem{White1996}
S.~R.~White, I.~Affleck, Phys.~Rev.~B {\bf 54}, 9862 (1996).

\bibitem{MajumdarGhosh}
C.~K.~Majumdar and D.~K.~Ghosh, J.~Math. Phys. {\bf 10} 1388 (1969);
C.~K.~Majumdar, J.~Phys. C {\bf 3}, 911 (1970).

\bibitem{ChandraDoucout}
P.~Chandra, B.~Doucot, Phys. Rev. B {\bf 38}, 9335 (1988).

\bibitem{Chakravarty1989}
S.~Chakravarty, B.~I.~Halperin, and D.~R.~Nelson, Phys. Rev. B {\bf 39}, 2344
(1989).

\bibitem{Lieb1999}
E.~H. Lieb and P.~Schupp, Phys. Rev. Lett. {\bf 83}, 5362 (1999).

\bibitem{Singh1999}
R.~R.~P.~Singh, Z.~Weihong, C.~J.~Hamer, and J.~Oitmaa, Phys. Rev. B {\bf 60},
7278 (1999).

\bibitem{Sushkov}
O.~P.~Sushkov, J.~Oitmaa, Z.~Weihong, Phys. Rev. B {\bf 66}, 054401 (2002);
\emph{ibid} {\bf 63}, 104420 (2001); V.~N.~Kotov, O.~P.~Sushkov, \emph{ibid}
{\bf 61}, 11820 (2000).

\bibitem{Tchernyshyov2003}
O.~Tchernyshyov, O.~A.~Starykh, R.~Moessner, A.~G.~Abanov, cond-mat/0301303
(2003).

\bibitem{Anderson1987}
P.~W.~Anderson, Science {\bf 235}, 1196 (1987).

\bibitem{Kivelson1987}
S.~A.~Kivelson, D.~S.~Rokhsar, and J.~P.~Sethna, Phys. Rev. B {\bf 35}, 8865
(1987).

\bibitem{Shender1982}
E.~Shender, Sov. Phys. JETP {\bf 56},178 (1982).

\bibitem{Henley1989}
C.~L.~Henley, Phys. Rev. Lett {\bf 62}, 2056 (1989); \emph{ibid} {\bf 73}, 2788
(1994).

\bibitem{Zaliznyak2000}
I.~A.~Zaliznyak, J.~P.~Hill, J.~M.~Tranquada, R.~Erwin, Y.~Moritomo,
Phys.~Rev.~Lett. {\bf 85}, 4353 (2000).

\bibitem{Tchernyshyov2002}
O.~Tchernyshyov, R.~Moessner and S.~L.~Sondhi, Phys. Rev. Lett. {\bf 88}, 067203
(2002) use same tag for a completely different phenomenon, arising from the
magnetoelastic coupling in the pyrochlore antiferromagnets.

\bibitem{Pytte}
E.~Pytte, Phys. Rev. B {\bf 10}, 4637 (1974).

\end{thebibliography}
\end{document}